\documentclass[%
 reprint,
 superscriptaddress,
 amsmath,amssymb,
 aps,
]{revtex4-2}

\usepackage{graphicx}
\usepackage{dcolumn}
\usepackage{bm}
\usepackage{lipsum}
\usepackage{braket}
\usepackage{siunitx}
\usepackage{epigraph} 
\usepackage{todonotes}
\setcitestyle{super} 

\usepackage{hyperref}

\begin{document}

\preprint{APS/123-QED}

\title{Gatemon Qubit Revisited for Improved Reliability and Stability }

\author{David Feldstein-Bofill}
\author{Zhenhai Sun}
\author{Casper Wied}
\author{Shikhar Singh}
\affiliation{Center for Quantum Devices, Niels Bohr Institute, University of Copenhagen, Denmark}
\affiliation{NNF Quantum Computing Programme, Niels Bohr Institute, University of Copenhagen, Denmark}

\author{Brian D. Isakov}
\affiliation{Department of Electrical, Computer \& Energy Engineering, University of Colorado Boulder, Boulder, CO 80309, USA}

\author{Svend Krøjer}
\affiliation{Center for Quantum Devices, Niels Bohr Institute, University of Copenhagen, Denmark}
\affiliation{NNF Quantum Computing Programme, Niels Bohr Institute, University of Copenhagen, Denmark}

\author{Jacob Hastrup}
\affiliation{Center for Quantum Devices, Niels Bohr Institute, University of Copenhagen, Denmark}
\affiliation{NNF Quantum Computing Programme, Niels Bohr Institute, University of Copenhagen, Denmark}

\author{Andr\'as Gyenis}
\affiliation{Department of Electrical, Computer \& Energy Engineering, University of Colorado Boulder, Boulder, CO 80309, USA}
\affiliation{Department of Physics, University of Colorado Boulder, Boulder CO 80309, USA}

\author{Morten Kjaergaard}
\affiliation{Center for Quantum Devices, Niels Bohr Institute, University of Copenhagen, Denmark}
\affiliation{NNF Quantum Computing Programme, Niels Bohr Institute, University of Copenhagen, Denmark}

\date{\today}

\begin{abstract}

The development of quantum circuits based on hybrid superconductor-semiconductor Josephson junctions holds promise for exploring their mesoscopic physics and for building novel superconducting devices. 
The gate-tunable superconducting transmon qubit (gatemon) is the paradigmatic example of such a superconducting circuit.
However, gatemons typically suffer from unstable and hysteretic qubit frequencies with respect to the applied gate voltage and reduced coherence times.
Here we develop methods for characterizing these challenges in gatemons and deploy these methods to compare the impact of shunt capacitor designs on gatemon performance. Our results indicate a strong frequency- and design-dependent behavior of the qubit stability, hysteresis, and dephasing times. Moreover, we achieve highly reliable tuning of the qubit frequency with $\SI{1}{MHz}$ precision over a range of several GHz, along with improved stability in grounded gatemons compared to gatemons with a floating capacitor design.
\end{abstract}

\maketitle

\section{\label{sec:level1}Introduction}

Josephson junctions are the foundational element of a wide range of quantum circuits, including superconducting qubits \cite{koch2007charge,krantz2019quantum}, parametric amplifiers \cite{yamamoto2008flux,esposito2021perspective}, quantum simulators \cite{carusotto2020photonic}, and quantum sensors \cite{sawicki2011sensitive,walsh2021josephson}. 
Most of these applications use a control knob which tunes the Josephson energy.
The common way to change the effective Josephson potential is by replacing a single junction with a superconducting loop containing two junctions and threading a magnetic field through the loop. In this loop, also referred to as a superconducting quantum interference device (SQUID) \cite{jaklevic1964quantum}, the effective Josephson energy depends on the magnetic flux due to the interference effects of the superconducting wavefunction.
This flux-tunability has played a central role in the development of superconducting qubits based on superconductor-insulator-superconductor (SIS) Josephson junctions.
In particular, the flexibility of flux-tunable transmon qubits \cite{koch2007charge} combined with their high coherence times \cite{wang2022towards,biznarova2024mitigation, somoroff2023millisecond,kjaergaard2020superconducting} have made them a promising platform for scalable quantum information systems. For example, flux-tunability enabled high fidelity two-qubit gates \cite{ding2023high,moskalenko2022high, sung2021realization}, flux-tunable resonators \cite{pierre2014storage}, fast unconditional reset of qubit states \cite{zhou2021rapid}, on-demand controllable dissipation \cite{maurya2024demand}, high-fidelity dispersive readout \cite{swiadek2023enhancing}, and addressing frequency crowding issues in larger multi-qubit systems \cite{versluis2017scalable}.

An alternative approach to realizing superconducting devices with tunable Josephson energies is to use hybrid superconductor-semiconductor-superconductor (S-Sm-S) Josephson junctions \cite{doh2005tunable}.
Here, an electrostatic voltage applied to a single hybrid junction controls the Josephson energy by tuning the number of Andreev bound states (ABSs) and their respective transmission.
Such hybrid junctions have been embedded in different types of superconducting qubits, for example, in the transmon \cite{larsen2015semiconductor}, the fluxonium \cite{strickland2024gatemonium, pita2020gate}, and in a protected qubit \cite{larsen2020parity}. 
Transmons based on hybrid junctions, named gatemons, have been realized, for example, using InAs nanowires with epitaxially grown superconducting layer \cite{larsen2015semiconductor,de2015realization}, proximitized two-dimensional electron gases\cite{casparis2018superconducting,sagi2024gate}, planar selective area grown InAs nanowires on silicon \cite{hertel2022gate},  planar Germanium\cite{Sagi2024} and graphene sheets\cite{wang2019coherent}. The development of semiconductor-based Josephson junctions has also led to novel superconducting circuits, including the Andreev spin qubit \cite{hays2021coherent}, gate-tunable resonators \cite{strickland2023superconducting}, and has opened new paths for current-phase relationship engineering \cite{banszerus2024hybrid}.

Despite their broad range of implementations and applications, these junctions have four general limitations: (1) unreliability of the qubit frequency vs. applied gate voltage, (2)  instability of the qubit frequency in time, (3) gate voltage sweep-direction-dependent hysteresis of the qubit frequency, and (4) reduced relaxation times compared to transmons.

\begin{figure}[t!]
\includegraphics{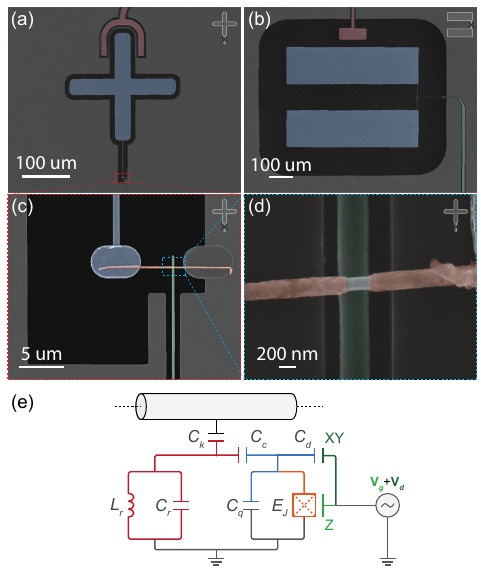}
\caption{\label{device} InAs/Al nanowire gatemon qubits. (a), (b) False-colored scanning electron microscope image of gatemons with grounded and floating capacitors (blue). A $\lambda/4$ resonator is capacitively coupled to the qubit for readout (red). (c),(d) Images of the nanowire junction. A segment of the epitaxial-coated aluminum (orange) is etched to form the S-Sm-S junction. The electrostatic gate (geeen) is placed underneath the etched region of the junction. The design has been optimized to reduce the capacitive coupling between the gate and the qubit island. (e) Circuit schematic comprising the feedline, readout resonator, gatemon qubit, and the drive line.}
\centering
\end{figure}

\begin{figure*}[t!]
\includegraphics{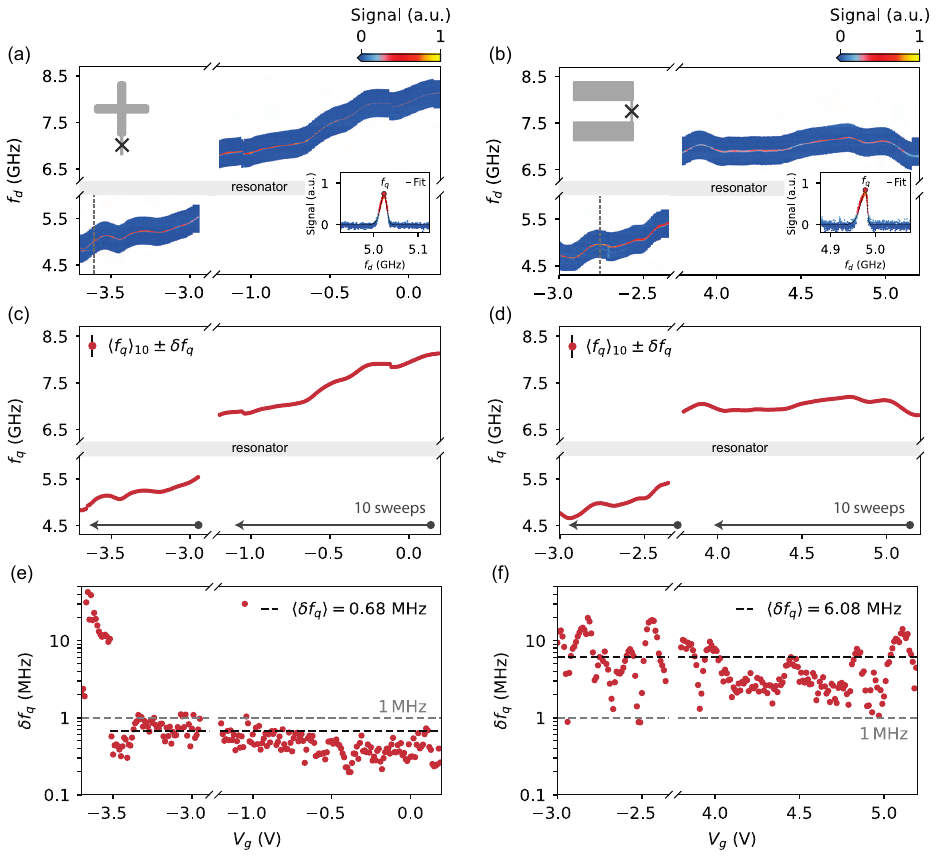}
\caption{\label{reliability}Reliability study of gatemons with grounded and floating designs. (a), (b) Two-tone spectroscopy as a function of the gate voltage of the two designs with colors indicating the magnitude of the transmission through the readout resonator. The measurement window for each trace is $\SI{300}{MHz}$. Inset figures show representative traces taken at the gate voltage marked with dark gray dashed lines. The same voltage sweep is repeated ten times. (c), (d) Average qubit frequency $\langle f_{q} \rangle$ and standard deviation $\delta f_q$ obtained from ten different voltage sweeps as a function of $V_{g}$. The qubit frequency $f_{q}$ is extracted through a Lorentzian/asymmetric Gaussian fit of the data. The arrows indicate the sweeping direction. (e), (f) Standard deviation of the qubit frequency across the ten sweeps. The black dashed lines show the mean of the standard deviation $\langle \delta f_{q} \rangle$ in the high-frequency regime for the grounded design and in the entire regime for the floating design. The gray dashed lines indicate $\SI{1}{MHz}$ for reference.}
\end{figure*}

\begin{figure*}[t!]
\includegraphics{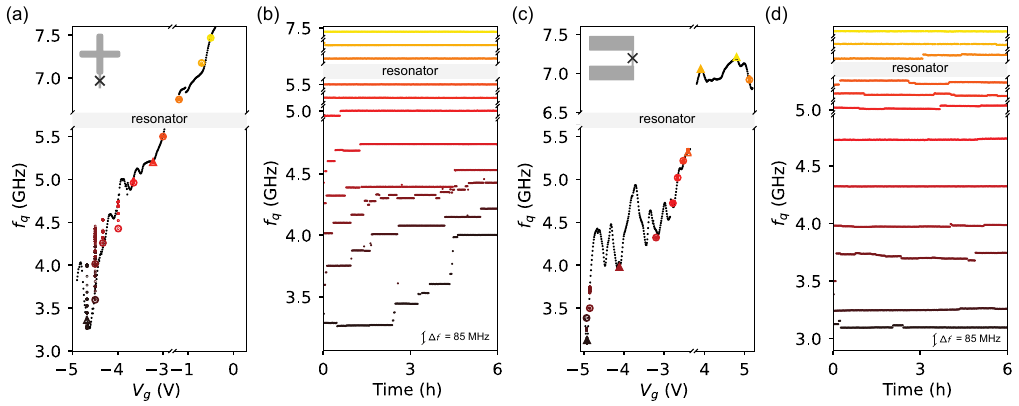}
\caption{\label{stability} Stability study of gatemons with grounded and floating designs. (a), (c) The qubit frequency $f_{q}$ (black solid markers) as a function of gate voltage $V_{g}$ for the two designs. The qubit is tuned to different gate values, including both gate sweet-spots ($\partial f_{q}/\partial V_g \approx 0$) and gate-sensitive points ($\partial f_{q}/\partial V_g \neq 0$). The circle/triangle markers indicate the initially measured qubit frequencies in the case of a gate voltage corresponding to a slope/sweet spot. At each gate voltage, the qubit is monitored for six hours. (b), (d) The extracted qubit frequency measured during the time window with colors corresponding to the data points in (a),(c). The window size for frequencies above $\SI{5}{GHz}$ is fixed to $\SI{85}{MHz}$.}
\centering
\end{figure*}

The aim of this study is to develop robust methods for addressing and improving these four issues.
As the platform for this work, we choose the gatemon qubit due to its relatively straightforward design, fabrication, and well-studied device parameters.
Here, we use InAs nanowires with epitaxially grown Al on all facets as the base for semiconductor junctions (see Supplement Material for fabrication details). 
We study two different geometries of the gatemon: 
a \textit{grounded} design [Fig.~\ref{device}(a)], and a \textit{floating} design [Fig.~\ref{device}(b)]. 
While both designs have similar charging energies (c.f. Table~\ref{tab:qubit_params}), they differ in the geometry of the capacitor pads. 
In the grounded design, the capacitor island is directly connected to the ground through the junction, providing a well-defined reference potential.
On the other hand, the floating design features capacitor pads that are separated from the ground plane. 
For both designs, the gate line supplies the DC gate voltage, $V_g$ to tune the junction energy, and provides the AC drive signal, $V_d$ to control the state of the qubit through capacitive coupling.
We carefully designed the gate line to balance the appropriate coupling without limiting qubit lifetimes due to spontaneous emission \cite{zhenhaisun2025}.

This paper is organized as follows: in Section~\ref{sec:reliability} we study the reliability of the qubit frequency with respect to gate voltages. 
In Section~\ref{sec:stability} we investigate the time-stability of grounded and floating gatemon devices. 
Section~\ref{sec:hysteresis} elucidates the role of hysteresis for gatemons. 
Finally, in Section~\ref{sec:coherence} we compare the coherence performance of grounded and floating gatemon designs.
The data presented in the main text corresponds to the same single set of grounded and floating qubits. Additional measurements on these two gatemons and data on two more devices from the same chip can be found in the Supplemental Sections.

\section{reliability}\label{sec:reliability}

The principles of the electrostatic and magnetic control of Josephson junctions are fundamentally different. As we briefly explain below, in the former, the intrinsic microscopic properties of the junctions influence the tunability, while in the latter, a macroscopic interference effect secures a reliably tunable Josephson energy.  

In gate-tunable transmons, the qubit frequency depends on the applied gate voltage as $f_{q}(V_g) \approx \sqrt{8 E_J(V_g) E_C}-E_C$, where  $E_J(V_g)$ is the voltage-dependent Josephson energy of the junction and $E_C$ is the charging energy. 
The physical mechanism behind voltage tunability is that the gate voltage modifies the density of states at the semiconductor part of the junction. This changes the number of channels allowed to carry supercurrent and the transmission probabilities of the Andreev states. 
The Josephson potential $U_{\text{S-Sm-S}}(V_g)$ takes the form \cite{beenakker1991universal}

\begin{equation}
    U_{\text{S-Sm-S}}(V_g) = - \Delta \sum_i \sqrt{1 - T_i(V_g) \sin^2{(\phi/2)}},
\end{equation}
where $\Delta$ is the proximitized superconducting gap, and the $T_i(V_g)$ transmission probabilities directly depend on the microscopic properties of the junction and the gate-voltage.
These probabilities change depending on the details of fabrication and the electrostatic environment near the junction.

On the other hand, in the case of flux-tunable transmon qubits, the qubit frequency depends on the external flux threading the loop of the device, $\Phi_{\text{ext}}$ as $f_{q}(\Phi_{\text{ext}}) \approx \sqrt{8 E_J(\Phi_{\text{ext}}) E_C} -E_C$\cite{koch2007charge}. Here, the Josephson energy changes with flux as $E_{J}(\Phi_{\text{ext}}) = 2 E_J | \cos(\Phi_{\text{ext}}) |$ \cite{tinkham2004introduction,krantz2019quantum} for symmetric junctions with Josephson energies $E_J$.
Since this effect relies on the interference of the macroscopic superconducting wavefunction, the external flux is a highly reliable control knob. 
The main deviation from this equation is the presence of spurious two-level systems \cite{andersen2017ultrastrong} and flux noise \cite{rower2023evolution}.
This means that for a given value of external flux, the variation in qubit frequencies upon repeated flux-sweeps is predominantly small, and the tuning of the qubit frequency is reliable.

To quantitatively assess the reliability of the gate-tunable transmon, we perform repeated spectroscopic measurements as a function of the gate voltage and record the qubit frequency $f_q(V_g)$.
At each gate voltage, we first determine the readout frequency and then apply a continuous microwave drive tone through the gate line to record the qubit frequency.
We repeat the entire gate voltage sweep ten times back-to-back.

Figures \ref{reliability}(a) and (b) show examples of the qubit spectroscopy dataset for the grounded and floating designs. 
We extract the qubit frequency via an asymmetric Gaussian fit \cite{azzalini1996multivariate} for frequencies below the resonator, and Lorentzian fit for frequencies above the resonator, as shown in the inset of Fig. \ref{reliability}(a), (b). 
We follow this procedure to account for the observed asymmetry in the spectroscopy of the qubit resonance. Such inhomogeneous broadening of the qubit peak can result from measurement conditions when the qubit frequency depends on the instantaneous value of the resonator photon number \cite{schuster2005ac}.

The mean qubit frequency $\langle f_q\rangle$ and its standard deviation $\delta f_q$ across the ten spectroscopy experiments for both grounded and floating designs are shown in Figs. \ref{reliability}(c), (d). 
The standard deviation error bars are not resolvable in the figure due to being smaller than their corresponding data points, hence we plot the standard deviation alone in Figs. \ref{reliability}(e), (f). 
For additional data on other devices see Appendix~\ref{reliability app}.

Overall, we find that the variation in qubit frequencies for the grounded design is an order of magnitude smaller than that of the floating design but with a stronger frequency-dependent behavior. 
For the grounded gatemon, the average standard deviation $\langle \delta f_{q} \rangle$ for qubit frequencies above $\SI{5.1}{GHz}$ is $\langle \delta f_{q} \rangle = \SI{0.68}{MHz}$, while for lower frequencies, the standard deviation increases. 
We note that there are two systematic charge jumps at $V_g= \SI{-1.02}{V}$ and $V_g= \SI{-0.1}{V}$ that abruptly change the qubit frequency by tens of megahertz. 
Similar charge jumps have been observed across the four devices, and their physical origin is not known.
The charge jumps at $V_g= \SI{-0.1}{V}$ are repeatable across the ten sweeps, while the jump at $V_g= \SI{-1.02}{V}$ is not repeatable, leading to a high standard deviation at that point.

For the floating gatemon, on the other hand, we find a standard deviation of $\langle \delta f_{q} \rangle = \SI{6.08}{MHz}$ across the measured frequency range, which is a factor of ten higher than the one observed in the grounded design at high frequencies. This variation, however, shows no strong frequency trend, unlike in the grounded design. The quantitative contrast of the standard deviation between the two gatemon geometries suggests that the grounded design is more reliable than the floating design at high frequencies. We hypothesize that the lack of galvanic connection to a well-defined ground in the floating gatemon potentially increases the qubit sensitivity to charge noise. 
Consequently, the qubit frequency can become more sensitive to fluctuations in the electric potential and lead to less reliable qubit frequency versus gate voltage sweeps.

\begin{figure}[t!]
\includegraphics{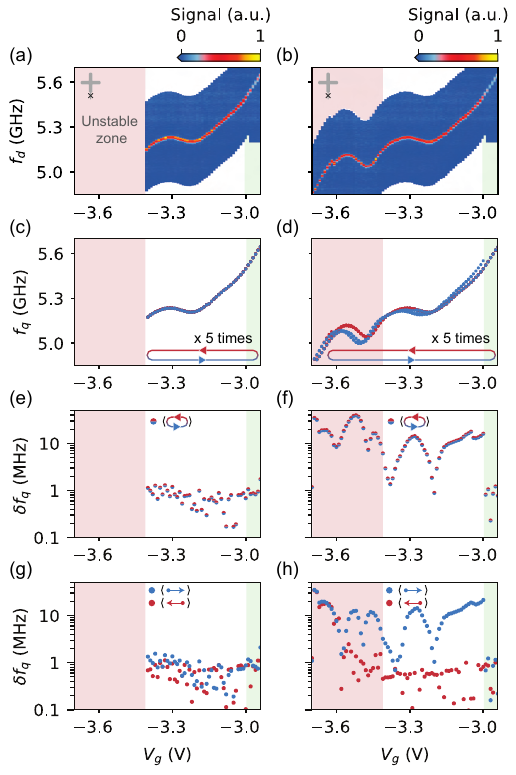}
\caption{\label{hysteresis} The hysteresis of qubit frequency as a function of gate voltage. (a), (b) Qubit spectroscopy as a function of the gate voltage for the grounded design. The voltage range in (a) is a subset of the voltage range in (b). The red shaded area shows the regime where the qubit frequency is unstable. The green shaded area shows the region where the qubit frequency points (both from upward and downward traces) are reliable. (c), (d) The extracted qubit frequencies across the ten different gate sweeps. The gate is swept downwards (red) and consecutively upwards (blue). The same procedure is repeated five times. (e), (f) Standard deviation of the qubit frequency $f_{q}$ of the ten traces. (g), (h) Standard deviation of downward traces (red) and upward traces (blue).}
\end{figure}

\section{stability}\label{sec:stability}

Previous gatemon studies have reported qubit frequency jumps and drifts over time \cite{luthi2018evolution,casparis2016gatemon}. In some cases, such instabilities have been associated with the qubit frequency sensitivity to the gate voltage. Building on these works we further develop an understanding of gatemon stability to consistently find the most stable operation regimes.

In this section, we characterize the frequency stability of both the grounded and the floating designs as a function of time across a wide frequency range. 
We choose various operating points in the non-monotonic gatemon spectrum, that not only have different frequencies but also different sensitivities to the gate voltage. 
We first set the gate voltage at a certain value and then monitor the qubit frequency through continuous-wave two-tone spectroscopy over six hours. 

Figures \ref{stability}(a) and (c) report the qubit frequency spectrum of the grounded and floating design as a function of $V_g$. 
Each large marker (triangle - sweet spot, circle - slope) indicates the initially measured qubit frequency at the applied gate voltage where we monitor the qubit frequency. 
Then, we plot in Figs.~\ref{stability}(b) and(d) the measured qubit frequency values as a function of time with small circles of the same color as the initial point. 
For additional data from separate devices see the Supplemental Material (c.f. Fig. \ref{stability app}).

In both designs, the qubits experience at least one of the two different instability behaviors: the qubit frequency jumps and/or continuously drifts over time. The behavior of the two designs is similar to the reliability measurements: the grounded design tends to have better performance at high frequencies and worse at low frequencies, while the instabilities are more consistent in the floating design.

For the grounded gatemon, the qubit frequency remains stable over the measured time in the high-frequency regime (above $\SI{5}{GHz}$ for this qubit). 
There are no observable discrete frequency jumps or drifts regardless of whether the qubit frequency is at or away from a sweet spot at these frequencies. 
For the floating gatemon, the qubit frequency drifts over time but does not show any discrete jumps when the qubit frequency is above $\SI{7}{GHz}$ in this device.
On the other hand, when the qubit frequency in the grounded design is at the low-frequency regime (below $\SI{5}{GHz}$ for this qubit), the qubit frequency exhibits significant discrete jumps, even at nominal sweet spots. 
For the lowest measured qubit frequency, the jumps result in an overall frequency change of around $\SI{1}{GHz}$ after 6 hours. 
In the floating design, the qubit frequency shows smaller discrete jumps as well as MHz-level drifting across the whole frequency range below $\SI{7}{GHz}$, both at a sweet spot or away from it.
These behaviors suggest that the qubit transition frequency is the dominant factor in the qubit stability, rather than its sensitivity to the gate voltage.

\section{Hysteresis}\label{sec:hysteresis}

In the flux-tunable transmon, the frequency depends on the exact value of the external flux, and, in the absence of magnetic impurities, it is independent of whether the external flux was swept down or up to reach the desired flux. 
Since the tunability of a semiconductor-based junction relies on the microscopic properties of the junction, the qubit frequency can depend on the history of the applied gate voltage. 
Thus, the qubit frequency can differ depending on whether the gate voltage is swept up ($\rightarrow$) or down ($\leftarrow$) to reach the target voltage value $V_g^{*}$. 
This leads to a hysteresis in the gate voltage, such that $f_q(V_g^{*,\rightarrow}) \neq f_q(V_g^{*,\leftarrow})$ \cite{luthi2018evolution}.

To understand the role of hysteresis, we focus on a grounded gatemon device. 
We study two different frequency ranges, both starting at the same frequency, but ending at different values, leading to distinct qubit behavior.
Fig. \ref{hysteresis}(a) shows the qubit frequency when the trace ends inside a \textit{reliable} frequency zone, while Fig. \ref{hysteresis}(b) shows data where the frequency ends in an \textit{unstable} frequency zone. 
For both cases, we study the qubit frequency as a function of gate voltage by sweeping in both negative and positive directions.

First, we sweep the gate voltage down and up five times, back-to-back, as shown in Fig. \ref{hysteresis}(c) and (d). Then, we compare the qubit frequency corresponding to down sweeps (sweeping towards negative gate voltages) and up sweeps (sweeping towards positive gate voltages).
Figures \ref{hysteresis}(e) and (f) show the standard deviation of the qubit frequency across all sweeps.
To elucidate the role of sweep-direction, we also plot separately the standard deviation going down in voltage (red points) and up in voltage (blue points) [Fig. \ref{hysteresis}(g), (h)]. 
We observe that the up sweep standard deviation in Fig. \ref{hysteresis}(h) dominates the shape of the total standard deviation seen in Fig. \ref{hysteresis}(f).
On the other hand, the down sweep standard deviation in Fig. \ref{hysteresis}(h) stays below \SI{1}{MHz} down to about $f_q \approx \SI{5.1}{GHz}$, where the standard deviation starts increasing. 
Note that the same behavior was observed in Fig. \ref{reliability}(e). 
The higher reliability of the down sweeps compared to the up sweeps stems from the final gate voltage points in the up sweep remaining reliable across all up sweeps. 
We highlight with green colors the gate voltage range where the qubit frequency points have been reliable. 
If the last point of the upsweep is not reliable, as shown in Appendix Fig. \ref{hysteresis_app}(c), the down sweeps can begin at different qubit frequencies. 
Consequently, the down sweep will only be reliable if the starting qubit frequency is the same.

In the left column of panels in Fig.\ \ref{hysteresis}, we study the hysteresis of the qubit frequency in a smaller gate voltage range, avoiding the unstable frequency zone.  
The low magnitudes of the standard deviation of the sweeps indicate that both gate sweep directions are equally reliable if both the starting and end qubit frequency points are reliable. 
In this specific device, this translates to a reliable frequency range above $\gtrsim \SI{5}{GHz}$. The protocol presented here can be used to determine the presence of unstable and reliable zones in gatemon devices.

\begin{figure}[t!]
\includegraphics{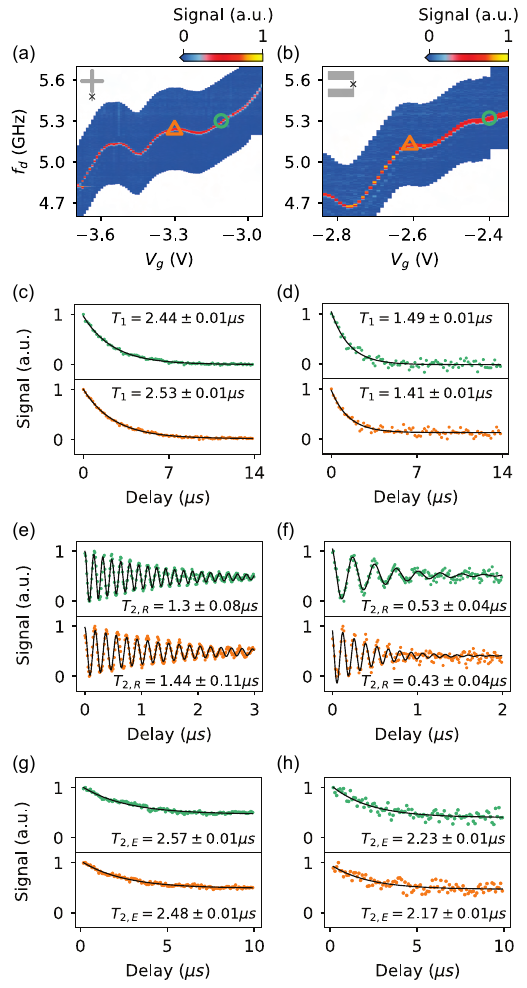}
\caption{\label{coherence} Coherence time study of gatemons with grounded and floating designs. (a), (b) Qubit two-tone spectroscopy showing the frequencies, where the coherence times are measured (circle: slope; triangle: sweet spot). (c), (d) Example for the relaxation time $T_{1}$ measurements for the two marked qubit frequencies. (e), (f) Ramsey interferometry for the two designs. The $T_{2,R}$ values are extracted by fitting the data to a damped sinusoidal function. Each trace is representative of a dataset where the experiment has been repeatedly measured over one hour. (g), (h) Example for the Hahn-Echo measurements on both designs. $T_{2,E}$ is extracted through an exponential decay fit.}
\end{figure}

\section{Coherence}\label{sec:coherence}

While the majority of gatemon studies focused on the grounded design with reported relaxation times reaching $\SI{10}{\mu s}$ \cite{larsen2015semiconductor, casparis2016gatemon,kringhoj2021magnetic,bargerbos2023mitigation,danilenko2023few}, the highest reported relaxation time ($\sim \SI{20}{\mu s}$) was measured in a floating design \cite{luthi2018evolution}. 
The difference in surface participation ratio of the electromagnetic fields \cite{wang2015surface} motivates the comparison of relaxation times of such designs on the same chip.

In this study, we measure energy relaxation time $T_1$ and dephasing time $T_2$ on grounded and floating gatemons fabricated on the same chip with readout resonators coupled to a common feedline. 
The latter experiment includes both low-frequency noise-sensitive Ramsey interferometry ($T_{2,R}$ dephasing time) and Hahn-Echo measurement ($T_{2,E}$ dephasing time) which is less sensitive to quasi-static noise. 
We study these coherence times at the first-order insensitive voltage sweet spot and away from it, as depicted in Fig. \ref{coherence}(a), (b). Additional experiments on other devices can be found in the Supplement. 
In all cases, we monitor the relaxation times for an hour, as shown in the Supplement (c.f. Fig. \ref{coh_app}, \ref{coherence app}).  
Furthermore, we choose similar frequencies for both designs to ensure a reliable comparison. 
The energy decay time $T_1$ ranges from $\approx \SI{2.4}{\mu s}$ at high qubit frequencies ($\approx \SI{5}{GHz}$) to $\approx \SI{8}{\mu s}$ at low qubit frequencies ($\approx \SI{3}{GHz}$). 
Figures \ref{coherence}(c) and (d) show examples of the measured relaxation curves for grounded and floating gatemon. 
Given that the two designs have significantly different electric field participation ratios, the similar relaxation rates indicate that the limiting loss may not be the dielectric loss associated with impurities on the surface of the capacitor pads but arises from a S-Sm-S junction-specific loss mechanism. 

Regarding the dephasing times, the grounded gatemon exhibits an average Ramsey dephasing time of $T_{2, R} \sim \SI{1.4}{\mu s}$, while the floating design shows a shorter $T_{2,R} \sim \SI{0.5}{\mu s}$ [Fig. \ref{coherence}(e), (f)]. 
No appreciable difference is observed for the relaxation and dephasing times when the qubit frequency is measured at a sweet spot or at a first-order charge-sensitive voltage value. Finally, both designs show Hahn-Echo dephasing times of $T_{2,E} \sim \SI{2}{\mu s}$ [Fig. \ref{coherence}(g), (h)]. 
Since both designs show similar $T_{2,E}$, while the floating gatemon has lower $T_{2,R}$, we can conclude that the floating gatemon design is more sensitive to low-frequency noise than the grounded design. 
This observation is in agreement with the drifting frequency behavior discussed in Sec. \ref{sec:stability}.

\section{Conclusions}

In summary, we have studied four limitations that affect the majority of gatemon experiments to date: reliability, stability, hysteresis, and decoherence. 
We have shown that by using a grounded design we can achieve qubit frequency reliability with precision below $\SI{1}{MHz}$ even when sweeping over a few GHz wide regions. 
Moreover, we demonstrated that gate sweep-direction-dependent qubit frequency modulation can be minimized by ensuring that both the start and end points of operations are within particularly reliable frequency regions.
Additionally, we find that the grounded design shows increased qubit frequency stability compared to a floating design at higher qubit frequencies. 
Finally, we found that the qubit frequency stability is most sensitive to the frequency itself rather than its derivative with respect to the gate voltage. 
We hypothesize that lower qubit frequencies, when the Josephson energy is smaller, may entail a smaller number of Andreev channels in the junction. The reduced number of channels carrying supercurrent across the junction could make the critical current more sensitive to noise fluctuations.
Despite no significant difference in the relaxation time of both qubit designs, we observed three times higher Ramsey dephasing times in the grounded qubit compared to the floating qubit at the gate voltage sweet spot. 
This finding and the fact that the Hahn-Echo times are similar in both designs suggest that the floating gatemon is more sensitive to low-frequency noise than its grounded counterpart.

Our work paves the way toward understanding, calibrating, and building more advanced hybrid superconducting-semiconducting circuits that are stable over time and can be operated reliably.

\begin{acknowledgments}
We thank Charlie Marcus for many useful discussions. We gratefully acknowledge Peter Krogstrup for growth of the nanowires. 
We also acknowledge the DTU cleanroom facilities and engineers for the support on nanofabrication.

We gratefully acknowledge support U.S. Army Research Office Grant No. W911NF-22-1-0042, the Novo Nordisk Foundation, Grant number NNF22SA0081175, the NNF Quantum Computing Programme, Villum Foundation (grant 37467) through a Villum Young Investigator grant, and the European Union (ERC Starting Grant, NovADePro, 101077479). Any opinions, findings, conclusions or recommendations expressed in this material are those of the author(s) and do not necessarily reflect the views of Army Research Office or the US Government. Views and opinions expressed are those of the author(s) only and do not necessarily reflect those of the European Union or the European Research Council. Neither the European Union nor the granting authority can be held responsible for them. Finally, we gratefully acknowledge Lena Jacobsen for program management support.

\end{acknowledgments}

\appendix

\section{Fabrication}

Our gatemons are fabricated on a $\SI{70}{nm}$ thick NbTiN superconducting layer sputtered on top of a high-resistivity silicon substrate. The device is patterned via dry reactive ion etching with $\mathrm{SF}_6/\mathrm{O}_2$ plasma. The InAs nanowire with epitaxial coated Al is deterministically placed on the contact pads with the use of a micromanipulator. Then, the SNS Josephson junction is formed by etching away $\sim\SI{200}{nm}$ long segment of a $\sim\SI{30}{nm}$ thick Al, epitaxially grown to all the InAs nanowire facets. The two ends of the nanowire are contacted to the control layer with Al patches with a previous Argon mill step to ensure good electrical contact.

\section{Devices}

The table below shows an overview of the devices on chip. 

\begin{table}[h!]
    \centering
    \begin{tabular}{||c c c c c||} 
     \hline
     Qubit & $f_R (\text{GHz})$ & $E_C (\text{MHz})$ & $C_{R,Q} (\text{fF})$ & $C_C (\text{fF})$ \\ [0.5ex] 
     \hline\hline
     Grounded & 6.12 & 220 & 4.92 & 0.155 \\ 
     \hline
     Grounded & 6.51 & 220 & 4.92 & 0.155 \\
     \hline
     Floating & 5.98 & 261 & 6.28 & 0.05 \\
     \hline
     Floating & 5.79 & 261 & 6.28 & 0.05 \\
     \hline
    \end{tabular}
    \caption{Summary of qubit parameters showing 1) Qubit design, 2) Measured readout resonator frequency, 3) Simulated qubit charging energy, 4) Simulated qubit-resonator coupling capacitance, 5) Simulated qubit-gate coupling capacitance.}
    \label{tab:qubit_params}
\end{table}

\begin{figure}[b!]
\includegraphics{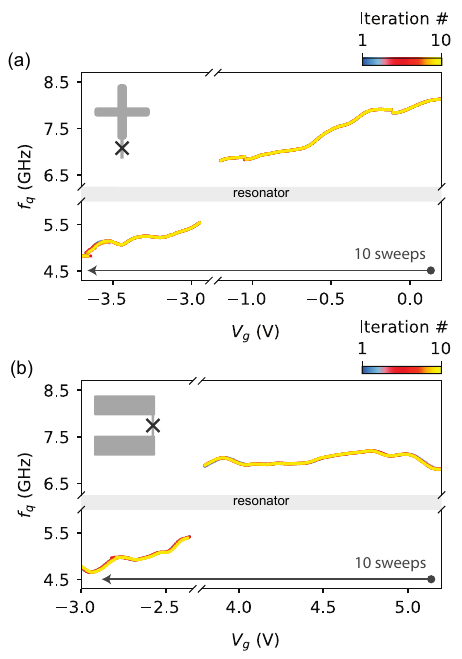}
\caption{\label{repr app} Reliability for grounded design and floating design. The 10 measured gatemon $f_{q}$ frequency traces for (a) grounded and (b) floating design shown in Figure \ref{reliability}. The color of each trace indicates the iteration number.}
\end{figure}

\begin{figure*}[t!]
\includegraphics{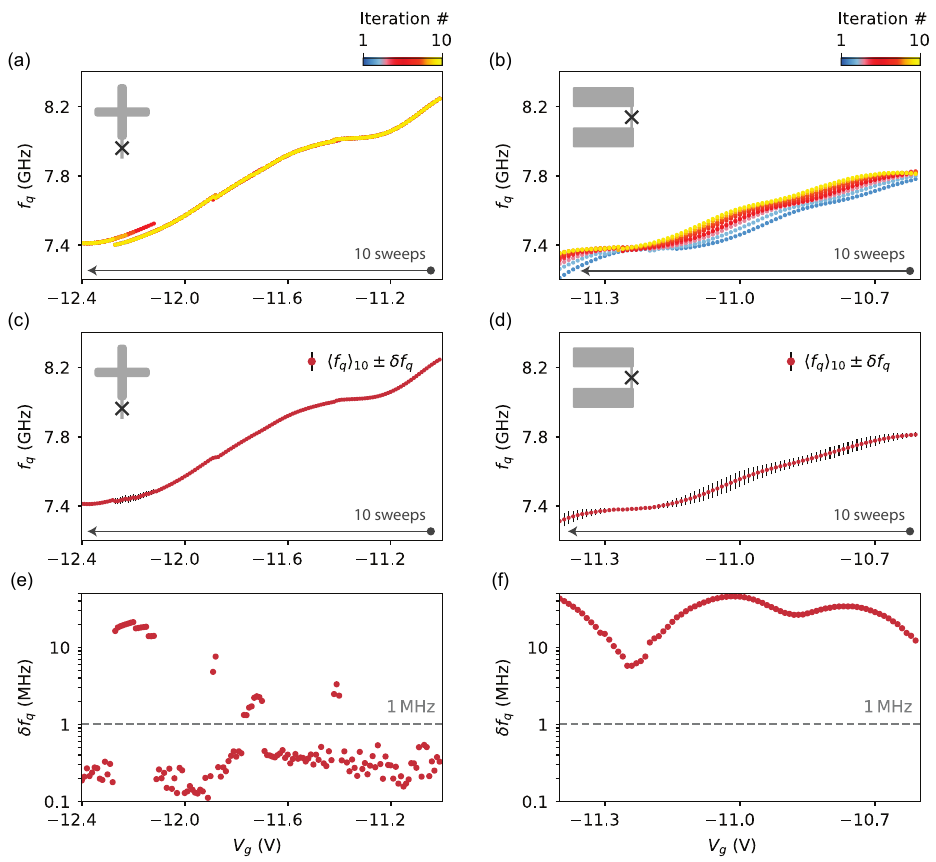}
\caption{\label{reliability app}Reliability for grounded design and floating design. The 10 measured gatemon $f_{q}$ traces for (a) grounded and (b) floating design. The color of each trace indicates the iteration number. (c),(d) Average and standard deviation of qubit frequency $f_{q}$ from ten different sweeps as a function of $V_{g}$. The qubit frequency $f_{q}$ is extracted through a Lorentzian fit of the data. The arrow indicates the sweeping direction. (e),(f) Standard deviation of $f_{q}$ across the ten sweeps. Grey dashed line placed at $\SI{1}{MHz}$ for reference.}
\end{figure*}

\section{Reliability}

This appendix provides supplementary measurements on the two gatemons presented in the main text plus data on two more devices from the same chip. 

Figure \ref{repr app} shows the ten traces used to study the reliability in the main text plotted separately. For the grounded gatemon the qubit frequency targetting becomes less reliable for lower frequencies. On the other hand, the floating gatemon shows similar unreliability across the studied frequency range.

Figure \ref{reliability app} includes the reliability study on a set of two additional gatemons from the same chip. Figures \ref{reliability app}(a),(b) show the ten gate voltage sweeps performed to study the reliability of the additional grounded and floating gatemon. Figures \ref{reliability app}(c),(d) show the average and standard deviation of the qubit frequency across the ten gate voltage sweeps. Finally, Figures \ref{reliability app}(e),(f) show the standard deviation across the ten sweeps. For the grounded gatemon the standard deviation stays below $\SI{1}{MHz}$ across the whole gate voltage range except near the observed charge jumps. Depending on the sweep number the charge jump occurs earlier or later in gate voltage, this effect can be seen in Figure \ref{reliability app}(a). For the floating gatemon, the standard deviation stays above $\SI{4}{MHz}$ across the studied gate voltage range. In this gatemon this occurs due to a shift of the qubit frequency towards higher frequency, as can be seen in Figure \ref{reliability app}(b).

The same reliability experiment has been done with non overlapping qubit spectroscopy measurements where no asymmetry was observed in the qubit peak. No quantitative difference was observed between the two experiments.

\begin{figure*}[t!]
\includegraphics{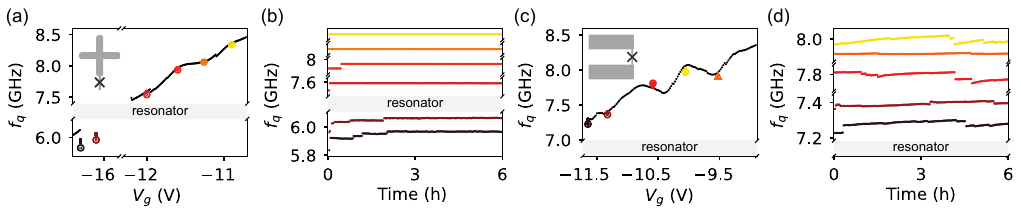}
\caption{\label{stability app}Stability for grounded design and floating design. Qubit frequency $f_{q}$ as a function of gate voltage $V_{g}$ for grounded (a) and floating (b) design shown in black solid markers. The qubit is tuned to different gate spots, including both gate sweet-spot ($\partial f_{q}/\partial V_g \approx 0$) and slope ($\partial f_{q}/\partial V_g \neq 0$). The big circle/triangle marker indicate the first measured qubit frequency point in the case of a slope/sweet-spot. At each gate voltage, the qubit is monitored for 6 hours. The qubit frequency measured during the whole time window is extracted and plotted in (b),(d). The data points are mapped to (a)(c) with corresponding colors.}
\end{figure*}

\begin{figure}[t!]
\includegraphics{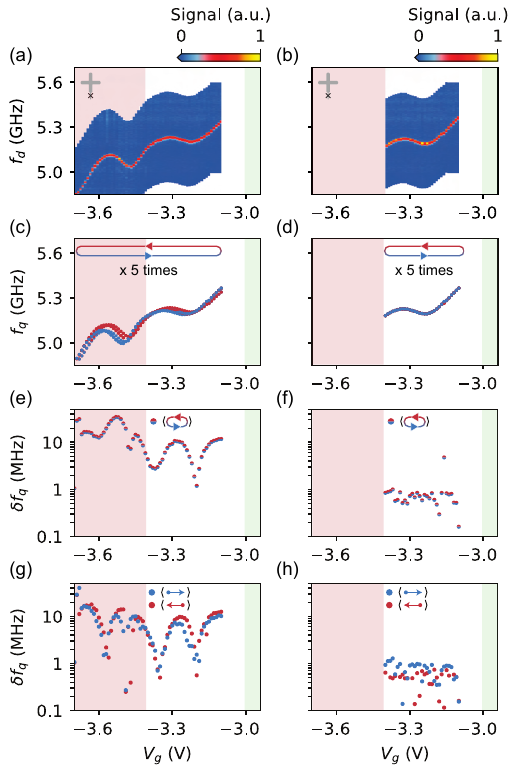}
\caption{\label{hysteresis_app} The hysteresis of qubit frequency as a function of gate voltage. (a),(b) Qubit spectroscopy as a function of the gate voltage for the grounded design. The voltage range in (b) is a subset of the voltage range in (a). The red shaded area shows the regime where the qubit frequency is unstable. The green shaded area shows the region where all the qubit frequency points (both from upward and downward traces) are reliable in Fig. \ref{hysteresis}. (c),(d) Extracted qubit frequency $f_{q}$ across the ten different gate sweeps. The gate is swept downwards (red) and consecutively upwards (blue). The same procedure is repeated five times. (e),(f) Standard deviation of the qubit frequency $f_{q}$ of the ten traces (g),(h) Standard deviation of downward traces (red) and upward traces (blue).}
\end{figure}

\section{Stability}

Figures \ref{stability app}(a) and (c) show the qubit frequency spectrum of an additional grounded and floating design as a function of $V_g$. 
Each large marker (triangle - sweet spot, circle - slope) indicates the initial measured qubit frequency at the gate voltage where we monitor the qubit frequency. 
Then, we plot the measured qubit frequency values as a function of time with small circles of the same color as the initial point [see  Figs.~\ref{stability app}(b),(d) with corresponding colors]. 

\begin{figure*}[h!]
\includegraphics{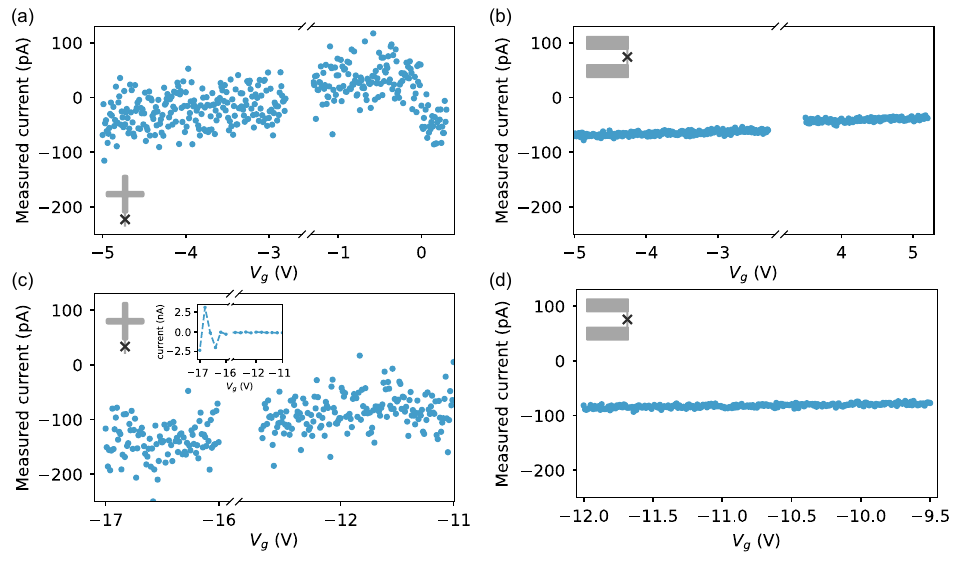}
\caption{\label{leak current app} Current measured across the studied gate voltage range for (a),(c) grounded design and (b),(d) floating design. Each data point is the mean of 10 measured data points at that gate voltage. Inset in (c) corresponds to an additional measurement taken on the same device on a different experiment.}
\end{figure*}

For the grounded gatemon, the qubit frequency remains stable over the measured time when $f_q$ is above \SI{8}{GHz}, regardless of the qubit frequency sensitivity to gate voltage at the measured gate spot. For the floating gatemon, the qubit frequency exhibits discrete jumps and drifts over time across all the studied frequency range. In contrast, the grounded qubit frequency shows discrete jumps below \SI{8}{GHz} combined with a mild drifting below \SI{6}{GHz}. Below \SI{6}{GHz}, corresponding to a gate voltage below \SI{-16}{V}, the gate starts slightly leaking current in the nA range, see inset in Fig. \ref{leak current app}(c). Hence such leaking current could be related to the mild frequency drift observed in this gate voltage range of this grounded gatemon.

\section{Hysteresis}

The data presented in this section is supplemental to the one shown in Figure \ref{hysteresis}. Here we show two additional frequency ranges, both starting at the same frequency, but ending at different values, resulting in different qubit frequency reliability.

Fig. \ref{hysteresis_app}(a) shows the qubit frequency trace where the end point occurs at an \textit{unstable} frequency zone, while Fig. \ref{hysteresis_app}(b) shows a trace where the frequency ends at a \textit{reliable} frequency zone. The main difference between this frequency range and the one presented in Fig. \ref{hysteresis} is the starting point. In Fig. \ref{hysteresis_app}(a),(b) the starting point lies away from the green shaded area, which is the range where all frequency points, regardless of gate sweep direction, were reliable.

For both frequency ranges, we sweep the gate voltage in both negative and positive directions five times, back-to-back, as shown in Fig. \ref{hysteresis_app}(c),(d). In Fig. \ref{hysteresis_app}(e),(f) we plot the standard deviation of the ten sweeps, while in Fig. \ref{hysteresis}(g),(h) we plot separately the standard deviation going down in voltage (red points) and up in voltage (blue points).

We observe a qualitatively similar down sweep and up sweep standard deviation in Fig. \ref{hysteresis_app}(g), in contrast with Fig. \ref{hysteresis}(e). That is because the final gate voltage point in the up sweeps has not been reliable. If the last point of the upsweep is not reliable, each down sweep begins at a different qubit frequency, making the qubit frequency curve with gate voltage different. On the other hand, both down sweeps and up sweeps' standard deviation stays below \SI{1}{MHz} in Fig. \ref{hysteresis_app}(h), showcasing that when both start and end point of the sweeps are inside the reliable zone hysteresis is minimized.

\section{Coherence}

\begin{figure}[t!]
\includegraphics{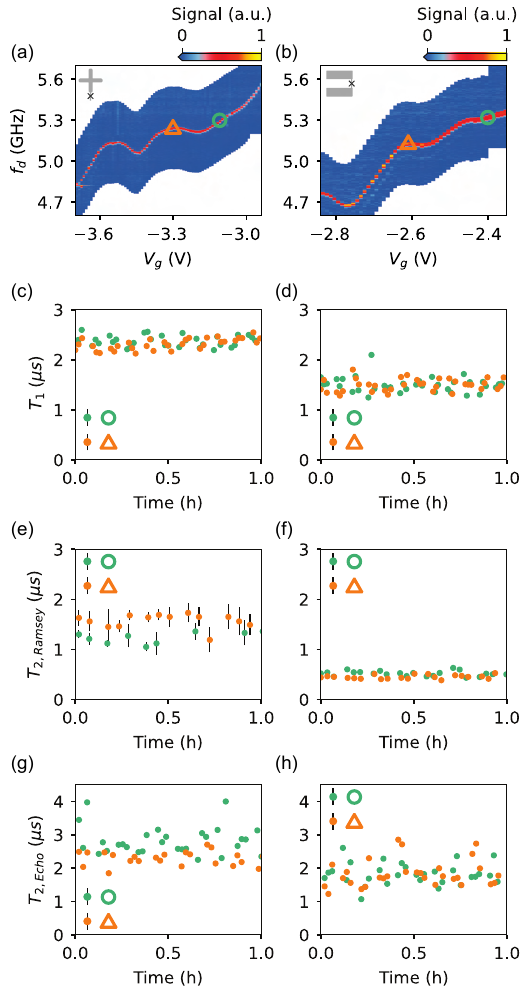}
\caption{\label{coh_app} Coherence time for grounded and floating gatemons. (a),(b) Qubit two-tone spectroscopy showing the frequencies, slope (circle) and sweet-spot (triangle), where the coherence times have been measured. (c),(d) Relaxation time $T_{1}$ for the two marked qubit frequencies monitored over 1 hour. (e),(f) Ramsey interferometry for both designs monitored over 1 hour. The $T_{2,ramsey}$ is extracted by fitting the data to a damped sinusoidal function. (g),(h) Hahn-Echo sequence for both designs monitored over 1 hour.}
\end{figure}

\begin{figure*}[t!]
\includegraphics{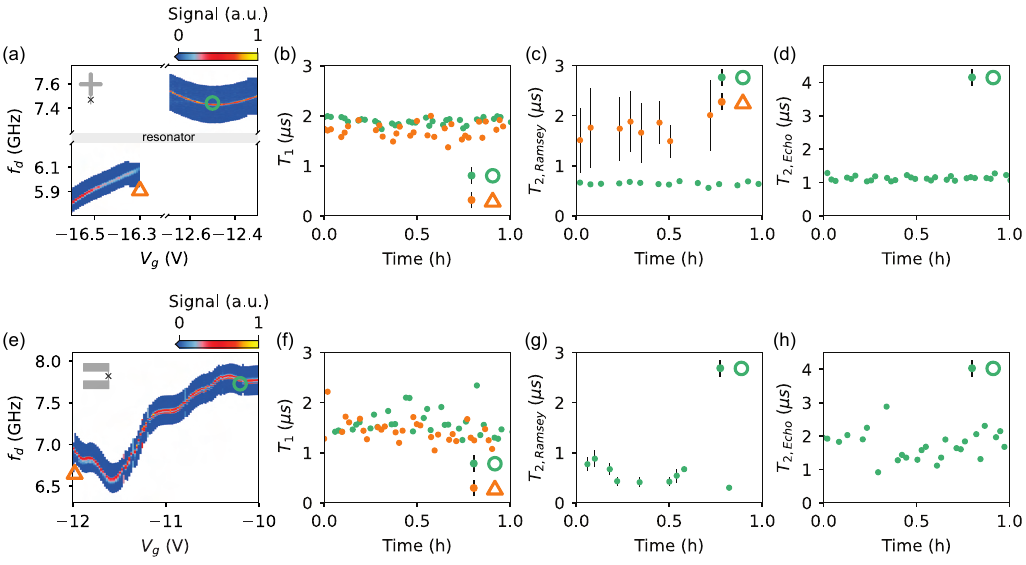}
\caption{\label{coherence app} Coherence time for grounded and floating gatemons. (a),(e) Qubit two-tone spectroscopy showing the frequencies, slope (circle) and sweet-spot (triangle), where the coherence times have been measured. (b),(f) Relaxation time $T_{1}$ for the two marked qubit frequencies. (c),(g) Ramsey interferometry for both designs. Data from the triangle marker not shown due to qubit's frequency significant drifting. (d),(h) Hahn-Echo sequence for both designs. Data from the triangle marker not shown due to qubit's frequency jump (d) and drift (h).}
\end{figure*}

Figure \ref{coh_app} reports additional data on coherence times of the grounded and floating gatemons presented in the main text. As described in Section \ref{sec:coherence}, relaxation and dephasing times are monitored over 1 hour. Figure \ref{coherence app} reports relaxation and dephasing times corresponding to the additional set of gatemons from the same chip.

\clearpage

\bibliography{references}

\begin{thebibliography}{47}%
\makeatletter
\providecommand \@ifxundefined [1]{%
 \@ifx{#1\undefined}
}%
\providecommand \@ifnum [1]{%
 \ifnum #1\expandafter \@firstoftwo
 \else \expandafter \@secondoftwo
 \fi
}%
\providecommand \@ifx [1]{%
 \ifx #1\expandafter \@firstoftwo
 \else \expandafter \@secondoftwo
 \fi
}%
\providecommand \natexlab [1]{#1}%
\providecommand \enquote  [1]{``#1''}%
\providecommand \bibnamefont  [1]{#1}%
\providecommand \bibfnamefont [1]{#1}%
\providecommand \citenamefont [1]{#1}%
\providecommand \href@noop [0]{\@secondoftwo}%
\providecommand \href [0]{\begingroup \@sanitize@url \@href}%
\providecommand \@href[1]{\@@startlink{#1}\@@href}%
\providecommand \@@href[1]{\endgroup#1\@@endlink}%
\providecommand \@sanitize@url [0]{\catcode `\\12\catcode `\$12\catcode `\&12\catcode `\#12\catcode `\^12\catcode `\_12\catcode `\%12\relax}%
\providecommand \@@startlink[1]{}%
\providecommand \@@endlink[0]{}%
\providecommand \url  [0]{\begingroup\@sanitize@url \@url }%
\providecommand \@url [1]{\endgroup\@href {#1}{\urlprefix }}%
\providecommand \urlprefix  [0]{URL }%
\providecommand \Eprint [0]{\href }%
\providecommand \doibase [0]{https://doi.org/}%
\providecommand \selectlanguage [0]{\@gobble}%
\providecommand \bibinfo  [0]{\@secondoftwo}%
\providecommand \bibfield  [0]{\@secondoftwo}%
\providecommand \translation [1]{[#1]}%
\providecommand \BibitemOpen [0]{}%
\providecommand \bibitemStop [0]{}%
\providecommand \bibitemNoStop [0]{.\EOS\space}%
\providecommand \EOS [0]{\spacefactor3000\relax}%
\providecommand \BibitemShut  [1]{\csname bibitem#1\endcsname}%
\let\auto@bib@innerbib\@empty
\bibitem [{\citenamefont {Koch}\ \emph {et~al.}(2007)\citenamefont {Koch}, \citenamefont {Yu}, \citenamefont {Gambetta}, \citenamefont {Houck}, \citenamefont {Schuster}, \citenamefont {Majer}, \citenamefont {Blais}, \citenamefont {Devoret}, \citenamefont {Girvin},\ and\ \citenamefont {Schoelkopf}}]{koch2007charge}%
  \BibitemOpen
  \bibfield  {author} {\bibinfo {author} {\bibfnamefont {J.}~\bibnamefont {Koch}}, \bibinfo {author} {\bibfnamefont {T.~M.}\ \bibnamefont {Yu}}, \bibinfo {author} {\bibfnamefont {J.}~\bibnamefont {Gambetta}}, \bibinfo {author} {\bibfnamefont {A.~A.}\ \bibnamefont {Houck}}, \bibinfo {author} {\bibfnamefont {D.~I.}\ \bibnamefont {Schuster}}, \bibinfo {author} {\bibfnamefont {J.}~\bibnamefont {Majer}}, \bibinfo {author} {\bibfnamefont {A.}~\bibnamefont {Blais}}, \bibinfo {author} {\bibfnamefont {M.~H.}\ \bibnamefont {Devoret}}, \bibinfo {author} {\bibfnamefont {S.~M.}\ \bibnamefont {Girvin}},\ and\ \bibinfo {author} {\bibfnamefont {R.~J.}\ \bibnamefont {Schoelkopf}},\ }\bibfield  {title} {\bibinfo {title} {Charge-insensitive qubit design derived from the cooper pair box},\ }\href@noop {} {\bibfield  {journal} {\bibinfo  {journal} {Physical Review A—Atomic, Molecular, and Optical Physics}\ }\textbf {\bibinfo {volume} {76}},\ \bibinfo {pages} {042319} (\bibinfo {year} {2007})}\BibitemShut {NoStop}%
\bibitem [{\citenamefont {Krantz}\ \emph {et~al.}(2019)\citenamefont {Krantz}, \citenamefont {Kjaergaard}, \citenamefont {Yan}, \citenamefont {Orlando}, \citenamefont {Gustavsson},\ and\ \citenamefont {Oliver}}]{krantz2019quantum}%
  \BibitemOpen
  \bibfield  {author} {\bibinfo {author} {\bibfnamefont {P.}~\bibnamefont {Krantz}}, \bibinfo {author} {\bibfnamefont {M.}~\bibnamefont {Kjaergaard}}, \bibinfo {author} {\bibfnamefont {F.}~\bibnamefont {Yan}}, \bibinfo {author} {\bibfnamefont {T.~P.}\ \bibnamefont {Orlando}}, \bibinfo {author} {\bibfnamefont {S.}~\bibnamefont {Gustavsson}},\ and\ \bibinfo {author} {\bibfnamefont {W.~D.}\ \bibnamefont {Oliver}},\ }\bibfield  {title} {\bibinfo {title} {A quantum engineer's guide to superconducting qubits},\ }\href@noop {} {\bibfield  {journal} {\bibinfo  {journal} {Applied physics reviews}\ }\textbf {\bibinfo {volume} {6}} (\bibinfo {year} {2019})}\BibitemShut {NoStop}%
\bibitem [{\citenamefont {Yamamoto}\ \emph {et~al.}(2008)\citenamefont {Yamamoto}, \citenamefont {Inomata}, \citenamefont {Watanabe}, \citenamefont {Matsuba}, \citenamefont {Miyazaki}, \citenamefont {Oliver}, \citenamefont {Nakamura},\ and\ \citenamefont {Tsai}}]{yamamoto2008flux}%
  \BibitemOpen
  \bibfield  {author} {\bibinfo {author} {\bibfnamefont {T.}~\bibnamefont {Yamamoto}}, \bibinfo {author} {\bibfnamefont {K.}~\bibnamefont {Inomata}}, \bibinfo {author} {\bibfnamefont {M.}~\bibnamefont {Watanabe}}, \bibinfo {author} {\bibfnamefont {K.}~\bibnamefont {Matsuba}}, \bibinfo {author} {\bibfnamefont {T.}~\bibnamefont {Miyazaki}}, \bibinfo {author} {\bibfnamefont {W.~D.}\ \bibnamefont {Oliver}}, \bibinfo {author} {\bibfnamefont {Y.}~\bibnamefont {Nakamura}},\ and\ \bibinfo {author} {\bibfnamefont {J.}~\bibnamefont {Tsai}},\ }\bibfield  {title} {\bibinfo {title} {Flux-driven josephson parametric amplifier},\ }\href@noop {} {\bibfield  {journal} {\bibinfo  {journal} {Applied Physics Letters}\ }\textbf {\bibinfo {volume} {93}} (\bibinfo {year} {2008})}\BibitemShut {NoStop}%
\bibitem [{\citenamefont {Esposito}\ \emph {et~al.}(2021)\citenamefont {Esposito}, \citenamefont {Ranadive}, \citenamefont {Planat},\ and\ \citenamefont {Roch}}]{esposito2021perspective}%
  \BibitemOpen
  \bibfield  {author} {\bibinfo {author} {\bibfnamefont {M.}~\bibnamefont {Esposito}}, \bibinfo {author} {\bibfnamefont {A.}~\bibnamefont {Ranadive}}, \bibinfo {author} {\bibfnamefont {L.}~\bibnamefont {Planat}},\ and\ \bibinfo {author} {\bibfnamefont {N.}~\bibnamefont {Roch}},\ }\bibfield  {title} {\bibinfo {title} {Perspective on traveling wave microwave parametric amplifiers},\ }\href@noop {} {\bibfield  {journal} {\bibinfo  {journal} {Applied Physics Letters}\ }\textbf {\bibinfo {volume} {119}} (\bibinfo {year} {2021})}\BibitemShut {NoStop}%
\bibitem [{\citenamefont {Carusotto}\ \emph {et~al.}(2020)\citenamefont {Carusotto}, \citenamefont {Houck}, \citenamefont {Koll{\'a}r}, \citenamefont {Roushan}, \citenamefont {Schuster},\ and\ \citenamefont {Simon}}]{carusotto2020photonic}%
  \BibitemOpen
  \bibfield  {author} {\bibinfo {author} {\bibfnamefont {I.}~\bibnamefont {Carusotto}}, \bibinfo {author} {\bibfnamefont {A.~A.}\ \bibnamefont {Houck}}, \bibinfo {author} {\bibfnamefont {A.~J.}\ \bibnamefont {Koll{\'a}r}}, \bibinfo {author} {\bibfnamefont {P.}~\bibnamefont {Roushan}}, \bibinfo {author} {\bibfnamefont {D.~I.}\ \bibnamefont {Schuster}},\ and\ \bibinfo {author} {\bibfnamefont {J.}~\bibnamefont {Simon}},\ }\bibfield  {title} {\bibinfo {title} {Photonic materials in circuit quantum electrodynamics},\ }\href@noop {} {\bibfield  {journal} {\bibinfo  {journal} {Nature Physics}\ }\textbf {\bibinfo {volume} {16}},\ \bibinfo {pages} {268} (\bibinfo {year} {2020})}\BibitemShut {NoStop}%
\bibitem [{\citenamefont {Sawicki}\ \emph {et~al.}(2011)\citenamefont {Sawicki}, \citenamefont {Stefanowicz},\ and\ \citenamefont {Ney}}]{sawicki2011sensitive}%
  \BibitemOpen
  \bibfield  {author} {\bibinfo {author} {\bibfnamefont {M.}~\bibnamefont {Sawicki}}, \bibinfo {author} {\bibfnamefont {W.}~\bibnamefont {Stefanowicz}},\ and\ \bibinfo {author} {\bibfnamefont {A.}~\bibnamefont {Ney}},\ }\bibfield  {title} {\bibinfo {title} {Sensitive squid magnetometry for studying nanomagnetism},\ }\href@noop {} {\bibfield  {journal} {\bibinfo  {journal} {Semiconductor Science and Technology}\ }\textbf {\bibinfo {volume} {26}},\ \bibinfo {pages} {064006} (\bibinfo {year} {2011})}\BibitemShut {NoStop}%
\bibitem [{\citenamefont {Walsh}\ \emph {et~al.}(2021)\citenamefont {Walsh}, \citenamefont {Jung}, \citenamefont {Lee}, \citenamefont {Efetov}, \citenamefont {Wu}, \citenamefont {Huang}, \citenamefont {Ohki}, \citenamefont {Taniguchi}, \citenamefont {Watanabe}, \citenamefont {Kim} \emph {et~al.}}]{walsh2021josephson}%
  \BibitemOpen
  \bibfield  {author} {\bibinfo {author} {\bibfnamefont {E.~D.}\ \bibnamefont {Walsh}}, \bibinfo {author} {\bibfnamefont {W.}~\bibnamefont {Jung}}, \bibinfo {author} {\bibfnamefont {G.-H.}\ \bibnamefont {Lee}}, \bibinfo {author} {\bibfnamefont {D.~K.}\ \bibnamefont {Efetov}}, \bibinfo {author} {\bibfnamefont {B.-I.}\ \bibnamefont {Wu}}, \bibinfo {author} {\bibfnamefont {K.-F.}\ \bibnamefont {Huang}}, \bibinfo {author} {\bibfnamefont {T.~A.}\ \bibnamefont {Ohki}}, \bibinfo {author} {\bibfnamefont {T.}~\bibnamefont {Taniguchi}}, \bibinfo {author} {\bibfnamefont {K.}~\bibnamefont {Watanabe}}, \bibinfo {author} {\bibfnamefont {P.}~\bibnamefont {Kim}}, \emph {et~al.},\ }\bibfield  {title} {\bibinfo {title} {Josephson junction infrared single-photon detector},\ }\href@noop {} {\bibfield  {journal} {\bibinfo  {journal} {Science}\ }\textbf {\bibinfo {volume} {372}},\ \bibinfo {pages} {409} (\bibinfo {year} {2021})}\BibitemShut {NoStop}%
\bibitem [{\citenamefont {Jaklevic}\ \emph {et~al.}(1964)\citenamefont {Jaklevic}, \citenamefont {Lambe}, \citenamefont {Silver},\ and\ \citenamefont {Mercereau}}]{jaklevic1964quantum}%
  \BibitemOpen
  \bibfield  {author} {\bibinfo {author} {\bibfnamefont {R.}~\bibnamefont {Jaklevic}}, \bibinfo {author} {\bibfnamefont {J.}~\bibnamefont {Lambe}}, \bibinfo {author} {\bibfnamefont {A.}~\bibnamefont {Silver}},\ and\ \bibinfo {author} {\bibfnamefont {J.}~\bibnamefont {Mercereau}},\ }\bibfield  {title} {\bibinfo {title} {Quantum interference effects in josephson tunneling},\ }\href@noop {} {\bibfield  {journal} {\bibinfo  {journal} {Physical Review Letters}\ }\textbf {\bibinfo {volume} {12}},\ \bibinfo {pages} {159} (\bibinfo {year} {1964})}\BibitemShut {NoStop}%
\bibitem [{\citenamefont {Wang}\ \emph {et~al.}(2022)\citenamefont {Wang}, \citenamefont {Li}, \citenamefont {Xu}, \citenamefont {Li}, \citenamefont {Wang}, \citenamefont {Yang}, \citenamefont {Mi}, \citenamefont {Liang}, \citenamefont {Su}, \citenamefont {Yang} \emph {et~al.}}]{wang2022towards}%
  \BibitemOpen
  \bibfield  {author} {\bibinfo {author} {\bibfnamefont {C.}~\bibnamefont {Wang}}, \bibinfo {author} {\bibfnamefont {X.}~\bibnamefont {Li}}, \bibinfo {author} {\bibfnamefont {H.}~\bibnamefont {Xu}}, \bibinfo {author} {\bibfnamefont {Z.}~\bibnamefont {Li}}, \bibinfo {author} {\bibfnamefont {J.}~\bibnamefont {Wang}}, \bibinfo {author} {\bibfnamefont {Z.}~\bibnamefont {Yang}}, \bibinfo {author} {\bibfnamefont {Z.}~\bibnamefont {Mi}}, \bibinfo {author} {\bibfnamefont {X.}~\bibnamefont {Liang}}, \bibinfo {author} {\bibfnamefont {T.}~\bibnamefont {Su}}, \bibinfo {author} {\bibfnamefont {C.}~\bibnamefont {Yang}}, \emph {et~al.},\ }\bibfield  {title} {\bibinfo {title} {Towards practical quantum computers: Transmon qubit with a lifetime approaching 0.5 milliseconds},\ }\href@noop {} {\bibfield  {journal} {\bibinfo  {journal} {npj Quantum Information}\ }\textbf {\bibinfo {volume} {8}},\ \bibinfo {pages} {3} (\bibinfo {year} {2022})}\BibitemShut {NoStop}%
\bibitem [{\citenamefont {Bizn{\'a}rov{\'a}}\ \emph {et~al.}(2024)\citenamefont {Bizn{\'a}rov{\'a}}, \citenamefont {Osman}, \citenamefont {Rehnman}, \citenamefont {Chayanun}, \citenamefont {Kri{\v{z}}an}, \citenamefont {Malmberg}, \citenamefont {Rommel}, \citenamefont {Warren}, \citenamefont {Delsing}, \citenamefont {Yurgens} \emph {et~al.}}]{biznarova2024mitigation}%
  \BibitemOpen
  \bibfield  {author} {\bibinfo {author} {\bibfnamefont {J.}~\bibnamefont {Bizn{\'a}rov{\'a}}}, \bibinfo {author} {\bibfnamefont {A.}~\bibnamefont {Osman}}, \bibinfo {author} {\bibfnamefont {E.}~\bibnamefont {Rehnman}}, \bibinfo {author} {\bibfnamefont {L.}~\bibnamefont {Chayanun}}, \bibinfo {author} {\bibfnamefont {C.}~\bibnamefont {Kri{\v{z}}an}}, \bibinfo {author} {\bibfnamefont {P.}~\bibnamefont {Malmberg}}, \bibinfo {author} {\bibfnamefont {M.}~\bibnamefont {Rommel}}, \bibinfo {author} {\bibfnamefont {C.}~\bibnamefont {Warren}}, \bibinfo {author} {\bibfnamefont {P.}~\bibnamefont {Delsing}}, \bibinfo {author} {\bibfnamefont {A.}~\bibnamefont {Yurgens}}, \emph {et~al.},\ }\bibfield  {title} {\bibinfo {title} {Mitigation of interfacial dielectric loss in aluminum-on-silicon superconducting qubits},\ }\href@noop {} {\bibfield  {journal} {\bibinfo  {journal} {npj Quantum Information}\ }\textbf {\bibinfo {volume} {10}},\ \bibinfo {pages} {78} (\bibinfo {year} {2024})}\BibitemShut {NoStop}%
\bibitem [{\citenamefont {Somoroff}\ \emph {et~al.}(2023)\citenamefont {Somoroff}, \citenamefont {Ficheux}, \citenamefont {Mencia}, \citenamefont {Xiong}, \citenamefont {Kuzmin},\ and\ \citenamefont {Manucharyan}}]{somoroff2023millisecond}%
  \BibitemOpen
  \bibfield  {author} {\bibinfo {author} {\bibfnamefont {A.}~\bibnamefont {Somoroff}}, \bibinfo {author} {\bibfnamefont {Q.}~\bibnamefont {Ficheux}}, \bibinfo {author} {\bibfnamefont {R.~A.}\ \bibnamefont {Mencia}}, \bibinfo {author} {\bibfnamefont {H.}~\bibnamefont {Xiong}}, \bibinfo {author} {\bibfnamefont {R.}~\bibnamefont {Kuzmin}},\ and\ \bibinfo {author} {\bibfnamefont {V.~E.}\ \bibnamefont {Manucharyan}},\ }\bibfield  {title} {\bibinfo {title} {Millisecond coherence in a superconducting qubit},\ }\href@noop {} {\bibfield  {journal} {\bibinfo  {journal} {Physical Review Letters}\ }\textbf {\bibinfo {volume} {130}},\ \bibinfo {pages} {267001} (\bibinfo {year} {2023})}\BibitemShut {NoStop}%
\bibitem [{\citenamefont {Kjaergaard}\ \emph {et~al.}(2020)\citenamefont {Kjaergaard}, \citenamefont {Schwartz}, \citenamefont {Braum{\"u}ller}, \citenamefont {Krantz}, \citenamefont {Wang}, \citenamefont {Gustavsson},\ and\ \citenamefont {Oliver}}]{kjaergaard2020superconducting}%
  \BibitemOpen
  \bibfield  {author} {\bibinfo {author} {\bibfnamefont {M.}~\bibnamefont {Kjaergaard}}, \bibinfo {author} {\bibfnamefont {M.~E.}\ \bibnamefont {Schwartz}}, \bibinfo {author} {\bibfnamefont {J.}~\bibnamefont {Braum{\"u}ller}}, \bibinfo {author} {\bibfnamefont {P.}~\bibnamefont {Krantz}}, \bibinfo {author} {\bibfnamefont {J.~I.-J.}\ \bibnamefont {Wang}}, \bibinfo {author} {\bibfnamefont {S.}~\bibnamefont {Gustavsson}},\ and\ \bibinfo {author} {\bibfnamefont {W.~D.}\ \bibnamefont {Oliver}},\ }\bibfield  {title} {\bibinfo {title} {Superconducting qubits: Current state of play},\ }\href@noop {} {\bibfield  {journal} {\bibinfo  {journal} {Annual Review of Condensed Matter Physics}\ }\textbf {\bibinfo {volume} {11}},\ \bibinfo {pages} {369} (\bibinfo {year} {2020})}\BibitemShut {NoStop}%
\bibitem [{\citenamefont {Ding}\ \emph {et~al.}(2023)\citenamefont {Ding}, \citenamefont {Hays}, \citenamefont {Sung}, \citenamefont {Kannan}, \citenamefont {An}, \citenamefont {Di~Paolo}, \citenamefont {Karamlou}, \citenamefont {Hazard}, \citenamefont {Azar}, \citenamefont {Kim} \emph {et~al.}}]{ding2023high}%
  \BibitemOpen
  \bibfield  {author} {\bibinfo {author} {\bibfnamefont {L.}~\bibnamefont {Ding}}, \bibinfo {author} {\bibfnamefont {M.}~\bibnamefont {Hays}}, \bibinfo {author} {\bibfnamefont {Y.}~\bibnamefont {Sung}}, \bibinfo {author} {\bibfnamefont {B.}~\bibnamefont {Kannan}}, \bibinfo {author} {\bibfnamefont {J.}~\bibnamefont {An}}, \bibinfo {author} {\bibfnamefont {A.}~\bibnamefont {Di~Paolo}}, \bibinfo {author} {\bibfnamefont {A.~H.}\ \bibnamefont {Karamlou}}, \bibinfo {author} {\bibfnamefont {T.~M.}\ \bibnamefont {Hazard}}, \bibinfo {author} {\bibfnamefont {K.}~\bibnamefont {Azar}}, \bibinfo {author} {\bibfnamefont {D.~K.}\ \bibnamefont {Kim}}, \emph {et~al.},\ }\bibfield  {title} {\bibinfo {title} {High-fidelity, frequency-flexible two-qubit fluxonium gates with a transmon coupler},\ }\href@noop {} {\bibfield  {journal} {\bibinfo  {journal} {Physical Review X}\ }\textbf {\bibinfo {volume} {13}},\ \bibinfo {pages} {031035} (\bibinfo {year} {2023})}\BibitemShut {NoStop}%
\bibitem [{\citenamefont {Moskalenko}\ \emph {et~al.}(2022)\citenamefont {Moskalenko}, \citenamefont {Simakov}, \citenamefont {Abramov}, \citenamefont {Grigorev}, \citenamefont {Moskalev}, \citenamefont {Pishchimova}, \citenamefont {Smirnov}, \citenamefont {Zikiy}, \citenamefont {Rodionov},\ and\ \citenamefont {Besedin}}]{moskalenko2022high}%
  \BibitemOpen
  \bibfield  {author} {\bibinfo {author} {\bibfnamefont {I.~N.}\ \bibnamefont {Moskalenko}}, \bibinfo {author} {\bibfnamefont {I.~A.}\ \bibnamefont {Simakov}}, \bibinfo {author} {\bibfnamefont {N.~N.}\ \bibnamefont {Abramov}}, \bibinfo {author} {\bibfnamefont {A.~A.}\ \bibnamefont {Grigorev}}, \bibinfo {author} {\bibfnamefont {D.~O.}\ \bibnamefont {Moskalev}}, \bibinfo {author} {\bibfnamefont {A.~A.}\ \bibnamefont {Pishchimova}}, \bibinfo {author} {\bibfnamefont {N.~S.}\ \bibnamefont {Smirnov}}, \bibinfo {author} {\bibfnamefont {E.~V.}\ \bibnamefont {Zikiy}}, \bibinfo {author} {\bibfnamefont {I.~A.}\ \bibnamefont {Rodionov}},\ and\ \bibinfo {author} {\bibfnamefont {I.~S.}\ \bibnamefont {Besedin}},\ }\bibfield  {title} {\bibinfo {title} {High fidelity two-qubit gates on fluxoniums using a tunable coupler},\ }\href@noop {} {\bibfield  {journal} {\bibinfo  {journal} {npj Quantum Information}\ }\textbf {\bibinfo {volume} {8}},\ \bibinfo {pages} {130} (\bibinfo {year} {2022})}\BibitemShut {NoStop}%
\bibitem [{\citenamefont {Sung}\ \emph {et~al.}(2021)\citenamefont {Sung}, \citenamefont {Ding}, \citenamefont {Braum{\"u}ller}, \citenamefont {Veps{\"a}l{\"a}inen}, \citenamefont {Kannan}, \citenamefont {Kjaergaard}, \citenamefont {Greene}, \citenamefont {Samach}, \citenamefont {McNally}, \citenamefont {Kim} \emph {et~al.}}]{sung2021realization}%
  \BibitemOpen
  \bibfield  {author} {\bibinfo {author} {\bibfnamefont {Y.}~\bibnamefont {Sung}}, \bibinfo {author} {\bibfnamefont {L.}~\bibnamefont {Ding}}, \bibinfo {author} {\bibfnamefont {J.}~\bibnamefont {Braum{\"u}ller}}, \bibinfo {author} {\bibfnamefont {A.}~\bibnamefont {Veps{\"a}l{\"a}inen}}, \bibinfo {author} {\bibfnamefont {B.}~\bibnamefont {Kannan}}, \bibinfo {author} {\bibfnamefont {M.}~\bibnamefont {Kjaergaard}}, \bibinfo {author} {\bibfnamefont {A.}~\bibnamefont {Greene}}, \bibinfo {author} {\bibfnamefont {G.~O.}\ \bibnamefont {Samach}}, \bibinfo {author} {\bibfnamefont {C.}~\bibnamefont {McNally}}, \bibinfo {author} {\bibfnamefont {D.}~\bibnamefont {Kim}}, \emph {et~al.},\ }\bibfield  {title} {\bibinfo {title} {Realization of high-fidelity cz and zz-free iswap gates with a tunable coupler},\ }\href@noop {} {\bibfield  {journal} {\bibinfo  {journal} {Physical Review X}\ }\textbf {\bibinfo {volume} {11}},\ \bibinfo {pages} {021058} (\bibinfo {year} {2021})}\BibitemShut {NoStop}%
\bibitem [{\citenamefont {Pierre}\ \emph {et~al.}(2014)\citenamefont {Pierre}, \citenamefont {Svensson}, \citenamefont {Raman~Sathyamoorthy}, \citenamefont {Johansson},\ and\ \citenamefont {Delsing}}]{pierre2014storage}%
  \BibitemOpen
  \bibfield  {author} {\bibinfo {author} {\bibfnamefont {M.}~\bibnamefont {Pierre}}, \bibinfo {author} {\bibfnamefont {I.-M.}\ \bibnamefont {Svensson}}, \bibinfo {author} {\bibfnamefont {S.}~\bibnamefont {Raman~Sathyamoorthy}}, \bibinfo {author} {\bibfnamefont {G.}~\bibnamefont {Johansson}},\ and\ \bibinfo {author} {\bibfnamefont {P.}~\bibnamefont {Delsing}},\ }\bibfield  {title} {\bibinfo {title} {Storage and on-demand release of microwaves using superconducting resonators with tunable coupling},\ }\href@noop {} {\bibfield  {journal} {\bibinfo  {journal} {Applied Physics Letters}\ }\textbf {\bibinfo {volume} {104}} (\bibinfo {year} {2014})}\BibitemShut {NoStop}%
\bibitem [{\citenamefont {Zhou}\ \emph {et~al.}(2021)\citenamefont {Zhou}, \citenamefont {Zhang}, \citenamefont {Yin}, \citenamefont {Huai}, \citenamefont {Gu}, \citenamefont {Xu}, \citenamefont {Allcock}, \citenamefont {Liu}, \citenamefont {Xi}, \citenamefont {Yu} \emph {et~al.}}]{zhou2021rapid}%
  \BibitemOpen
  \bibfield  {author} {\bibinfo {author} {\bibfnamefont {Y.}~\bibnamefont {Zhou}}, \bibinfo {author} {\bibfnamefont {Z.}~\bibnamefont {Zhang}}, \bibinfo {author} {\bibfnamefont {Z.}~\bibnamefont {Yin}}, \bibinfo {author} {\bibfnamefont {S.}~\bibnamefont {Huai}}, \bibinfo {author} {\bibfnamefont {X.}~\bibnamefont {Gu}}, \bibinfo {author} {\bibfnamefont {X.}~\bibnamefont {Xu}}, \bibinfo {author} {\bibfnamefont {J.}~\bibnamefont {Allcock}}, \bibinfo {author} {\bibfnamefont {F.}~\bibnamefont {Liu}}, \bibinfo {author} {\bibfnamefont {G.}~\bibnamefont {Xi}}, \bibinfo {author} {\bibfnamefont {Q.}~\bibnamefont {Yu}}, \emph {et~al.},\ }\bibfield  {title} {\bibinfo {title} {Rapid and unconditional parametric reset protocol for tunable superconducting qubits},\ }\href@noop {} {\bibfield  {journal} {\bibinfo  {journal} {Nature Communications}\ }\textbf {\bibinfo {volume} {12}},\ \bibinfo {pages} {5924} (\bibinfo {year} {2021})}\BibitemShut {NoStop}%
\bibitem [{\citenamefont {Maurya}\ \emph {et~al.}(2024)\citenamefont {Maurya}, \citenamefont {Zhang}, \citenamefont {Kowsari}, \citenamefont {Kuo}, \citenamefont {Hartsell}, \citenamefont {Miyamoto}, \citenamefont {Liu}, \citenamefont {Shanto}, \citenamefont {Vlachos}, \citenamefont {Zarassi} \emph {et~al.}}]{maurya2024demand}%
  \BibitemOpen
  \bibfield  {author} {\bibinfo {author} {\bibfnamefont {V.}~\bibnamefont {Maurya}}, \bibinfo {author} {\bibfnamefont {H.}~\bibnamefont {Zhang}}, \bibinfo {author} {\bibfnamefont {D.}~\bibnamefont {Kowsari}}, \bibinfo {author} {\bibfnamefont {A.}~\bibnamefont {Kuo}}, \bibinfo {author} {\bibfnamefont {D.~M.}\ \bibnamefont {Hartsell}}, \bibinfo {author} {\bibfnamefont {C.}~\bibnamefont {Miyamoto}}, \bibinfo {author} {\bibfnamefont {J.}~\bibnamefont {Liu}}, \bibinfo {author} {\bibfnamefont {S.}~\bibnamefont {Shanto}}, \bibinfo {author} {\bibfnamefont {E.}~\bibnamefont {Vlachos}}, \bibinfo {author} {\bibfnamefont {A.}~\bibnamefont {Zarassi}}, \emph {et~al.},\ }\bibfield  {title} {\bibinfo {title} {On-demand driven dissipation for cavity reset and cooling},\ }\href@noop {} {\bibfield  {journal} {\bibinfo  {journal} {PRX Quantum}\ }\textbf {\bibinfo {volume} {5}},\ \bibinfo {pages} {020321} (\bibinfo {year} {2024})}\BibitemShut {NoStop}%
\bibitem [{\citenamefont {Swiadek}\ \emph {et~al.}(2023)\citenamefont {Swiadek}, \citenamefont {Shillito}, \citenamefont {Magnard}, \citenamefont {Remm}, \citenamefont {Hellings}, \citenamefont {Lacroix}, \citenamefont {Ficheux}, \citenamefont {Zanuz}, \citenamefont {Norris}, \citenamefont {Blais} \emph {et~al.}}]{swiadek2023enhancing}%
  \BibitemOpen
  \bibfield  {author} {\bibinfo {author} {\bibfnamefont {F.}~\bibnamefont {Swiadek}}, \bibinfo {author} {\bibfnamefont {R.}~\bibnamefont {Shillito}}, \bibinfo {author} {\bibfnamefont {P.}~\bibnamefont {Magnard}}, \bibinfo {author} {\bibfnamefont {A.}~\bibnamefont {Remm}}, \bibinfo {author} {\bibfnamefont {C.}~\bibnamefont {Hellings}}, \bibinfo {author} {\bibfnamefont {N.}~\bibnamefont {Lacroix}}, \bibinfo {author} {\bibfnamefont {Q.}~\bibnamefont {Ficheux}}, \bibinfo {author} {\bibfnamefont {D.~C.}\ \bibnamefont {Zanuz}}, \bibinfo {author} {\bibfnamefont {G.~J.}\ \bibnamefont {Norris}}, \bibinfo {author} {\bibfnamefont {A.}~\bibnamefont {Blais}}, \emph {et~al.},\ }\bibfield  {title} {\bibinfo {title} {Enhancing dispersive readout of superconducting qubits through dynamic control of the dispersive shift: Experiment and theory},\ }\href@noop {} {\bibfield  {journal} {\bibinfo  {journal} {arXiv preprint arXiv:2307.07765}\ } (\bibinfo {year} {2023})}\BibitemShut {NoStop}%
\bibitem [{\citenamefont {Versluis}\ \emph {et~al.}(2017)\citenamefont {Versluis}, \citenamefont {Poletto}, \citenamefont {Khammassi}, \citenamefont {Tarasinski}, \citenamefont {Haider}, \citenamefont {Michalak}, \citenamefont {Bruno}, \citenamefont {Bertels},\ and\ \citenamefont {DiCarlo}}]{versluis2017scalable}%
  \BibitemOpen
  \bibfield  {author} {\bibinfo {author} {\bibfnamefont {R.}~\bibnamefont {Versluis}}, \bibinfo {author} {\bibfnamefont {S.}~\bibnamefont {Poletto}}, \bibinfo {author} {\bibfnamefont {N.}~\bibnamefont {Khammassi}}, \bibinfo {author} {\bibfnamefont {B.}~\bibnamefont {Tarasinski}}, \bibinfo {author} {\bibfnamefont {N.}~\bibnamefont {Haider}}, \bibinfo {author} {\bibfnamefont {D.~J.}\ \bibnamefont {Michalak}}, \bibinfo {author} {\bibfnamefont {A.}~\bibnamefont {Bruno}}, \bibinfo {author} {\bibfnamefont {K.}~\bibnamefont {Bertels}},\ and\ \bibinfo {author} {\bibfnamefont {L.}~\bibnamefont {DiCarlo}},\ }\bibfield  {title} {\bibinfo {title} {Scalable quantum circuit and control for a superconducting surface code},\ }\href@noop {} {\bibfield  {journal} {\bibinfo  {journal} {Physical Review Applied}\ }\textbf {\bibinfo {volume} {8}},\ \bibinfo {pages} {034021} (\bibinfo {year} {2017})}\BibitemShut {NoStop}%
\bibitem [{\citenamefont {Doh}\ \emph {et~al.}(2005)\citenamefont {Doh}, \citenamefont {van Dam}, \citenamefont {Roest}, \citenamefont {Bakkers}, \citenamefont {Kouwenhoven},\ and\ \citenamefont {De~Franceschi}}]{doh2005tunable}%
  \BibitemOpen
  \bibfield  {author} {\bibinfo {author} {\bibfnamefont {Y.-J.}\ \bibnamefont {Doh}}, \bibinfo {author} {\bibfnamefont {J.~A.}\ \bibnamefont {van Dam}}, \bibinfo {author} {\bibfnamefont {A.~L.}\ \bibnamefont {Roest}}, \bibinfo {author} {\bibfnamefont {E.~P.}\ \bibnamefont {Bakkers}}, \bibinfo {author} {\bibfnamefont {L.~P.}\ \bibnamefont {Kouwenhoven}},\ and\ \bibinfo {author} {\bibfnamefont {S.}~\bibnamefont {De~Franceschi}},\ }\bibfield  {title} {\bibinfo {title} {Tunable supercurrent through semiconductor nanowires},\ }\href@noop {} {\bibfield  {journal} {\bibinfo  {journal} {science}\ }\textbf {\bibinfo {volume} {309}},\ \bibinfo {pages} {272} (\bibinfo {year} {2005})}\BibitemShut {NoStop}%
\bibitem [{\citenamefont {Larsen}\ \emph {et~al.}(2015)\citenamefont {Larsen}, \citenamefont {Petersson}, \citenamefont {Kuemmeth}, \citenamefont {Jespersen}, \citenamefont {Krogstrup}, \citenamefont {Nyg{\aa}rd},\ and\ \citenamefont {Marcus}}]{larsen2015semiconductor}%
  \BibitemOpen
  \bibfield  {author} {\bibinfo {author} {\bibfnamefont {T.~W.}\ \bibnamefont {Larsen}}, \bibinfo {author} {\bibfnamefont {K.~D.}\ \bibnamefont {Petersson}}, \bibinfo {author} {\bibfnamefont {F.}~\bibnamefont {Kuemmeth}}, \bibinfo {author} {\bibfnamefont {T.~S.}\ \bibnamefont {Jespersen}}, \bibinfo {author} {\bibfnamefont {P.}~\bibnamefont {Krogstrup}}, \bibinfo {author} {\bibfnamefont {J.}~\bibnamefont {Nyg{\aa}rd}},\ and\ \bibinfo {author} {\bibfnamefont {C.~M.}\ \bibnamefont {Marcus}},\ }\bibfield  {title} {\bibinfo {title} {Semiconductor-nanowire-based superconducting qubit},\ }\href@noop {} {\bibfield  {journal} {\bibinfo  {journal} {Physical review letters}\ }\textbf {\bibinfo {volume} {115}},\ \bibinfo {pages} {127001} (\bibinfo {year} {2015})}\BibitemShut {NoStop}%
\bibitem [{\citenamefont {Strickland}\ \emph {et~al.}(2024)\citenamefont {Strickland}, \citenamefont {Elfeky}, \citenamefont {Baker}, \citenamefont {Maiani}, \citenamefont {Lee}, \citenamefont {Levy}, \citenamefont {Issokson}, \citenamefont {Vrajitoarea},\ and\ \citenamefont {Shabani}}]{strickland2024gatemonium}%
  \BibitemOpen
  \bibfield  {author} {\bibinfo {author} {\bibfnamefont {W.~M.}\ \bibnamefont {Strickland}}, \bibinfo {author} {\bibfnamefont {B.~H.}\ \bibnamefont {Elfeky}}, \bibinfo {author} {\bibfnamefont {L.}~\bibnamefont {Baker}}, \bibinfo {author} {\bibfnamefont {A.}~\bibnamefont {Maiani}}, \bibinfo {author} {\bibfnamefont {J.}~\bibnamefont {Lee}}, \bibinfo {author} {\bibfnamefont {I.}~\bibnamefont {Levy}}, \bibinfo {author} {\bibfnamefont {J.}~\bibnamefont {Issokson}}, \bibinfo {author} {\bibfnamefont {A.}~\bibnamefont {Vrajitoarea}},\ and\ \bibinfo {author} {\bibfnamefont {J.}~\bibnamefont {Shabani}},\ }\bibfield  {title} {\bibinfo {title} {Gatemonium: A voltage-tunable fluxonium},\ }\href@noop {} {\bibfield  {journal} {\bibinfo  {journal} {arXiv preprint arXiv:2406.09002}\ } (\bibinfo {year} {2024})}\BibitemShut {NoStop}%
\bibitem [{\citenamefont {Pita-Vidal}\ \emph {et~al.}(2020)\citenamefont {Pita-Vidal}, \citenamefont {Bargerbos}, \citenamefont {Yang}, \citenamefont {Van~Woerkom}, \citenamefont {Pfaff}, \citenamefont {Haider}, \citenamefont {Krogstrup}, \citenamefont {Kouwenhoven}, \citenamefont {De~Lange},\ and\ \citenamefont {Kou}}]{pita2020gate}%
  \BibitemOpen
  \bibfield  {author} {\bibinfo {author} {\bibfnamefont {M.}~\bibnamefont {Pita-Vidal}}, \bibinfo {author} {\bibfnamefont {A.}~\bibnamefont {Bargerbos}}, \bibinfo {author} {\bibfnamefont {C.-K.}\ \bibnamefont {Yang}}, \bibinfo {author} {\bibfnamefont {D.~J.}\ \bibnamefont {Van~Woerkom}}, \bibinfo {author} {\bibfnamefont {W.}~\bibnamefont {Pfaff}}, \bibinfo {author} {\bibfnamefont {N.}~\bibnamefont {Haider}}, \bibinfo {author} {\bibfnamefont {P.}~\bibnamefont {Krogstrup}}, \bibinfo {author} {\bibfnamefont {L.~P.}\ \bibnamefont {Kouwenhoven}}, \bibinfo {author} {\bibfnamefont {G.}~\bibnamefont {De~Lange}},\ and\ \bibinfo {author} {\bibfnamefont {A.}~\bibnamefont {Kou}},\ }\bibfield  {title} {\bibinfo {title} {Gate-tunable field-compatible fluxonium},\ }\href@noop {} {\bibfield  {journal} {\bibinfo  {journal} {Physical Review Applied}\ }\textbf {\bibinfo {volume} {14}},\ \bibinfo {pages} {064038} (\bibinfo {year} {2020})}\BibitemShut {NoStop}%
\bibitem [{\citenamefont {Larsen}\ \emph {et~al.}(2020)\citenamefont {Larsen}, \citenamefont {Gershenson}, \citenamefont {Casparis}, \citenamefont {Kringh{\o}j}, \citenamefont {Pearson}, \citenamefont {McNeil}, \citenamefont {Kuemmeth}, \citenamefont {Krogstrup}, \citenamefont {Petersson},\ and\ \citenamefont {Marcus}}]{larsen2020parity}%
  \BibitemOpen
  \bibfield  {author} {\bibinfo {author} {\bibfnamefont {T.~W.}\ \bibnamefont {Larsen}}, \bibinfo {author} {\bibfnamefont {M.~E.}\ \bibnamefont {Gershenson}}, \bibinfo {author} {\bibfnamefont {L.}~\bibnamefont {Casparis}}, \bibinfo {author} {\bibfnamefont {A.}~\bibnamefont {Kringh{\o}j}}, \bibinfo {author} {\bibfnamefont {N.~J.}\ \bibnamefont {Pearson}}, \bibinfo {author} {\bibfnamefont {R.~P.}\ \bibnamefont {McNeil}}, \bibinfo {author} {\bibfnamefont {F.}~\bibnamefont {Kuemmeth}}, \bibinfo {author} {\bibfnamefont {P.}~\bibnamefont {Krogstrup}}, \bibinfo {author} {\bibfnamefont {K.~D.}\ \bibnamefont {Petersson}},\ and\ \bibinfo {author} {\bibfnamefont {C.~M.}\ \bibnamefont {Marcus}},\ }\bibfield  {title} {\bibinfo {title} {Parity-protected superconductor-semiconductor qubit},\ }\href@noop {} {\bibfield  {journal} {\bibinfo  {journal} {Physical review letters}\ }\textbf {\bibinfo {volume} {125}},\ \bibinfo {pages} {056801} (\bibinfo {year} {2020})}\BibitemShut {NoStop}%
\bibitem [{\citenamefont {De~Lange}\ \emph {et~al.}(2015)\citenamefont {De~Lange}, \citenamefont {Van~Heck}, \citenamefont {Bruno}, \citenamefont {Van~Woerkom}, \citenamefont {Geresdi}, \citenamefont {Plissard}, \citenamefont {Bakkers}, \citenamefont {Akhmerov},\ and\ \citenamefont {DiCarlo}}]{de2015realization}%
  \BibitemOpen
  \bibfield  {author} {\bibinfo {author} {\bibfnamefont {G.}~\bibnamefont {De~Lange}}, \bibinfo {author} {\bibfnamefont {B.}~\bibnamefont {Van~Heck}}, \bibinfo {author} {\bibfnamefont {A.}~\bibnamefont {Bruno}}, \bibinfo {author} {\bibfnamefont {D.}~\bibnamefont {Van~Woerkom}}, \bibinfo {author} {\bibfnamefont {A.}~\bibnamefont {Geresdi}}, \bibinfo {author} {\bibfnamefont {S.}~\bibnamefont {Plissard}}, \bibinfo {author} {\bibfnamefont {E.}~\bibnamefont {Bakkers}}, \bibinfo {author} {\bibfnamefont {A.}~\bibnamefont {Akhmerov}},\ and\ \bibinfo {author} {\bibfnamefont {L.}~\bibnamefont {DiCarlo}},\ }\bibfield  {title} {\bibinfo {title} {Realization of microwave quantum circuits using hybrid superconducting-semiconducting nanowire josephson elements},\ }\href@noop {} {\bibfield  {journal} {\bibinfo  {journal} {Physical review letters}\ }\textbf {\bibinfo {volume} {115}},\ \bibinfo {pages} {127002} (\bibinfo {year} {2015})}\BibitemShut {NoStop}%
\bibitem [{\citenamefont {Casparis}\ \emph {et~al.}(2018)\citenamefont {Casparis}, \citenamefont {Connolly}, \citenamefont {Kjaergaard}, \citenamefont {Pearson}, \citenamefont {Kringh{\o}j}, \citenamefont {Larsen}, \citenamefont {Kuemmeth}, \citenamefont {Wang}, \citenamefont {Thomas}, \citenamefont {Gronin} \emph {et~al.}}]{casparis2018superconducting}%
  \BibitemOpen
  \bibfield  {author} {\bibinfo {author} {\bibfnamefont {L.}~\bibnamefont {Casparis}}, \bibinfo {author} {\bibfnamefont {M.~R.}\ \bibnamefont {Connolly}}, \bibinfo {author} {\bibfnamefont {M.}~\bibnamefont {Kjaergaard}}, \bibinfo {author} {\bibfnamefont {N.~J.}\ \bibnamefont {Pearson}}, \bibinfo {author} {\bibfnamefont {A.}~\bibnamefont {Kringh{\o}j}}, \bibinfo {author} {\bibfnamefont {T.~W.}\ \bibnamefont {Larsen}}, \bibinfo {author} {\bibfnamefont {F.}~\bibnamefont {Kuemmeth}}, \bibinfo {author} {\bibfnamefont {T.}~\bibnamefont {Wang}}, \bibinfo {author} {\bibfnamefont {C.}~\bibnamefont {Thomas}}, \bibinfo {author} {\bibfnamefont {S.}~\bibnamefont {Gronin}}, \emph {et~al.},\ }\bibfield  {title} {\bibinfo {title} {Superconducting gatemon qubit based on a proximitized two-dimensional electron gas},\ }\href@noop {} {\bibfield  {journal} {\bibinfo  {journal} {Nature nanotechnology}\ }\textbf {\bibinfo {volume} {13}},\ \bibinfo {pages} {915} (\bibinfo {year} {2018})}\BibitemShut {NoStop}%
\bibitem [{\citenamefont {Sagi}\ \emph {et~al.}(2024{\natexlab{a}})\citenamefont {Sagi}, \citenamefont {Crippa}, \citenamefont {Valentini}, \citenamefont {Janik}, \citenamefont {Baghumyan}, \citenamefont {Fabris}, \citenamefont {Kapoor}, \citenamefont {Hassani}, \citenamefont {Fink}, \citenamefont {Calcaterra} \emph {et~al.}}]{sagi2024gate}%
  \BibitemOpen
  \bibfield  {author} {\bibinfo {author} {\bibfnamefont {O.}~\bibnamefont {Sagi}}, \bibinfo {author} {\bibfnamefont {A.}~\bibnamefont {Crippa}}, \bibinfo {author} {\bibfnamefont {M.}~\bibnamefont {Valentini}}, \bibinfo {author} {\bibfnamefont {M.}~\bibnamefont {Janik}}, \bibinfo {author} {\bibfnamefont {L.}~\bibnamefont {Baghumyan}}, \bibinfo {author} {\bibfnamefont {G.}~\bibnamefont {Fabris}}, \bibinfo {author} {\bibfnamefont {L.}~\bibnamefont {Kapoor}}, \bibinfo {author} {\bibfnamefont {F.}~\bibnamefont {Hassani}}, \bibinfo {author} {\bibfnamefont {J.}~\bibnamefont {Fink}}, \bibinfo {author} {\bibfnamefont {S.}~\bibnamefont {Calcaterra}}, \emph {et~al.},\ }\bibfield  {title} {\bibinfo {title} {A gate tunable transmon qubit in planar ge},\ }\href@noop {} {\bibfield  {journal} {\bibinfo  {journal} {Nature Communications}\ }\textbf {\bibinfo {volume} {15}},\ \bibinfo {pages} {6400} (\bibinfo {year} {2024}{\natexlab{a}})}\BibitemShut {NoStop}%
\bibitem [{\citenamefont {Hertel}\ \emph {et~al.}(2022)\citenamefont {Hertel}, \citenamefont {Eichinger}, \citenamefont {Andersen}, \citenamefont {van Zanten}, \citenamefont {Kallatt}, \citenamefont {Scarlino}, \citenamefont {Kringh{\o}j}, \citenamefont {Chavez-Garcia}, \citenamefont {Gardner}, \citenamefont {Gronin} \emph {et~al.}}]{hertel2022gate}%
  \BibitemOpen
  \bibfield  {author} {\bibinfo {author} {\bibfnamefont {A.}~\bibnamefont {Hertel}}, \bibinfo {author} {\bibfnamefont {M.}~\bibnamefont {Eichinger}}, \bibinfo {author} {\bibfnamefont {L.~O.}\ \bibnamefont {Andersen}}, \bibinfo {author} {\bibfnamefont {D.~M.}\ \bibnamefont {van Zanten}}, \bibinfo {author} {\bibfnamefont {S.}~\bibnamefont {Kallatt}}, \bibinfo {author} {\bibfnamefont {P.}~\bibnamefont {Scarlino}}, \bibinfo {author} {\bibfnamefont {A.}~\bibnamefont {Kringh{\o}j}}, \bibinfo {author} {\bibfnamefont {J.~M.}\ \bibnamefont {Chavez-Garcia}}, \bibinfo {author} {\bibfnamefont {G.~C.}\ \bibnamefont {Gardner}}, \bibinfo {author} {\bibfnamefont {S.}~\bibnamefont {Gronin}}, \emph {et~al.},\ }\bibfield  {title} {\bibinfo {title} {Gate-tunable transmon using selective-area-grown superconductor-semiconductor hybrid structures on silicon},\ }\href@noop {} {\bibfield  {journal} {\bibinfo  {journal} {Physical Review Applied}\ }\textbf {\bibinfo {volume} {18}},\ \bibinfo {pages} {034042} (\bibinfo {year}
  {2022})}\BibitemShut {NoStop}%
\bibitem [{\citenamefont {Sagi}\ \emph {et~al.}(2024{\natexlab{b}})\citenamefont {Sagi}, \citenamefont {Crippa}, \citenamefont {Valentini}, \citenamefont {Janik}, \citenamefont {Baghumyan}, \citenamefont {Fabris}, \citenamefont {Kapoor}, \citenamefont {Hassani}, \citenamefont {Fink}, \citenamefont {Calcaterra}, \citenamefont {Chrastina}, \citenamefont {Isella},\ and\ \citenamefont {Katsaros}}]{Sagi2024}%
  \BibitemOpen
  \bibfield  {author} {\bibinfo {author} {\bibfnamefont {O.}~\bibnamefont {Sagi}}, \bibinfo {author} {\bibfnamefont {A.}~\bibnamefont {Crippa}}, \bibinfo {author} {\bibfnamefont {M.}~\bibnamefont {Valentini}}, \bibinfo {author} {\bibfnamefont {M.}~\bibnamefont {Janik}}, \bibinfo {author} {\bibfnamefont {L.}~\bibnamefont {Baghumyan}}, \bibinfo {author} {\bibfnamefont {G.}~\bibnamefont {Fabris}}, \bibinfo {author} {\bibfnamefont {L.}~\bibnamefont {Kapoor}}, \bibinfo {author} {\bibfnamefont {F.}~\bibnamefont {Hassani}}, \bibinfo {author} {\bibfnamefont {J.}~\bibnamefont {Fink}}, \bibinfo {author} {\bibfnamefont {S.}~\bibnamefont {Calcaterra}}, \bibinfo {author} {\bibfnamefont {D.}~\bibnamefont {Chrastina}}, \bibinfo {author} {\bibfnamefont {G.}~\bibnamefont {Isella}},\ and\ \bibinfo {author} {\bibfnamefont {G.}~\bibnamefont {Katsaros}},\ }\bibfield  {title} {\bibinfo {title} {A gate tunable transmon qubit in planar ge},\ }\href {https://doi.org/10.1038/s41467-024-50763-6} {\bibfield  {journal} {\bibinfo
  {journal} {Nature Communications}\ }\textbf {\bibinfo {volume} {15}},\ \bibinfo {pages} {6400} (\bibinfo {year} {2024}{\natexlab{b}})}\BibitemShut {NoStop}%
\bibitem [{\citenamefont {Wang}\ \emph {et~al.}(2019)\citenamefont {Wang}, \citenamefont {Rodan-Legrain}, \citenamefont {Bretheau}, \citenamefont {Campbell}, \citenamefont {Kannan}, \citenamefont {Kim}, \citenamefont {Kjaergaard}, \citenamefont {Krantz}, \citenamefont {Samach}, \citenamefont {Yan} \emph {et~al.}}]{wang2019coherent}%
  \BibitemOpen
  \bibfield  {author} {\bibinfo {author} {\bibfnamefont {J.~I.-J.}\ \bibnamefont {Wang}}, \bibinfo {author} {\bibfnamefont {D.}~\bibnamefont {Rodan-Legrain}}, \bibinfo {author} {\bibfnamefont {L.}~\bibnamefont {Bretheau}}, \bibinfo {author} {\bibfnamefont {D.~L.}\ \bibnamefont {Campbell}}, \bibinfo {author} {\bibfnamefont {B.}~\bibnamefont {Kannan}}, \bibinfo {author} {\bibfnamefont {D.}~\bibnamefont {Kim}}, \bibinfo {author} {\bibfnamefont {M.}~\bibnamefont {Kjaergaard}}, \bibinfo {author} {\bibfnamefont {P.}~\bibnamefont {Krantz}}, \bibinfo {author} {\bibfnamefont {G.~O.}\ \bibnamefont {Samach}}, \bibinfo {author} {\bibfnamefont {F.}~\bibnamefont {Yan}}, \emph {et~al.},\ }\bibfield  {title} {\bibinfo {title} {Coherent control of a hybrid superconducting circuit made with graphene-based van der waals heterostructures},\ }\href@noop {} {\bibfield  {journal} {\bibinfo  {journal} {Nature nanotechnology}\ }\textbf {\bibinfo {volume} {14}},\ \bibinfo {pages} {120} (\bibinfo {year} {2019})}\BibitemShut {NoStop}%
\bibitem [{\citenamefont {Hays}\ \emph {et~al.}(2021)\citenamefont {Hays}, \citenamefont {Fatemi}, \citenamefont {Bouman}, \citenamefont {Cerrillo}, \citenamefont {Diamond}, \citenamefont {Serniak}, \citenamefont {Connolly}, \citenamefont {Krogstrup}, \citenamefont {Nyg{\aa}rd}, \citenamefont {Levy~Yeyati} \emph {et~al.}}]{hays2021coherent}%
  \BibitemOpen
  \bibfield  {author} {\bibinfo {author} {\bibfnamefont {M.}~\bibnamefont {Hays}}, \bibinfo {author} {\bibfnamefont {V.}~\bibnamefont {Fatemi}}, \bibinfo {author} {\bibfnamefont {D.}~\bibnamefont {Bouman}}, \bibinfo {author} {\bibfnamefont {J.}~\bibnamefont {Cerrillo}}, \bibinfo {author} {\bibfnamefont {S.}~\bibnamefont {Diamond}}, \bibinfo {author} {\bibfnamefont {K.}~\bibnamefont {Serniak}}, \bibinfo {author} {\bibfnamefont {T.}~\bibnamefont {Connolly}}, \bibinfo {author} {\bibfnamefont {P.}~\bibnamefont {Krogstrup}}, \bibinfo {author} {\bibfnamefont {J.}~\bibnamefont {Nyg{\aa}rd}}, \bibinfo {author} {\bibfnamefont {A.}~\bibnamefont {Levy~Yeyati}}, \emph {et~al.},\ }\bibfield  {title} {\bibinfo {title} {Coherent manipulation of an andreev spin qubit},\ }\href@noop {} {\bibfield  {journal} {\bibinfo  {journal} {Science}\ }\textbf {\bibinfo {volume} {373}},\ \bibinfo {pages} {430} (\bibinfo {year} {2021})}\BibitemShut {NoStop}%
\bibitem [{\citenamefont {Strickland}\ \emph {et~al.}(2023)\citenamefont {Strickland}, \citenamefont {Elfeky}, \citenamefont {Yuan}, \citenamefont {Schiela}, \citenamefont {Yu}, \citenamefont {Langone}, \citenamefont {Vavilov}, \citenamefont {Manucharyan},\ and\ \citenamefont {Shabani}}]{strickland2023superconducting}%
  \BibitemOpen
  \bibfield  {author} {\bibinfo {author} {\bibfnamefont {W.~M.}\ \bibnamefont {Strickland}}, \bibinfo {author} {\bibfnamefont {B.~H.}\ \bibnamefont {Elfeky}}, \bibinfo {author} {\bibfnamefont {J.~O.}\ \bibnamefont {Yuan}}, \bibinfo {author} {\bibfnamefont {W.~F.}\ \bibnamefont {Schiela}}, \bibinfo {author} {\bibfnamefont {P.}~\bibnamefont {Yu}}, \bibinfo {author} {\bibfnamefont {D.}~\bibnamefont {Langone}}, \bibinfo {author} {\bibfnamefont {M.~G.}\ \bibnamefont {Vavilov}}, \bibinfo {author} {\bibfnamefont {V.~E.}\ \bibnamefont {Manucharyan}},\ and\ \bibinfo {author} {\bibfnamefont {J.}~\bibnamefont {Shabani}},\ }\bibfield  {title} {\bibinfo {title} {Superconducting resonators with voltage-controlled frequency and nonlinearity},\ }\href@noop {} {\bibfield  {journal} {\bibinfo  {journal} {Physical Review Applied}\ }\textbf {\bibinfo {volume} {19}},\ \bibinfo {pages} {034021} (\bibinfo {year} {2023})}\BibitemShut {NoStop}%
\bibitem [{\citenamefont {Banszerus}\ \emph {et~al.}(2024)\citenamefont {Banszerus}, \citenamefont {Andersson}, \citenamefont {Marshall}, \citenamefont {Lindemann}, \citenamefont {Manfra}, \citenamefont {Marcus},\ and\ \citenamefont {Vaitiek{\.e}nas}}]{banszerus2024hybrid}%
  \BibitemOpen
  \bibfield  {author} {\bibinfo {author} {\bibfnamefont {L.}~\bibnamefont {Banszerus}}, \bibinfo {author} {\bibfnamefont {C.}~\bibnamefont {Andersson}}, \bibinfo {author} {\bibfnamefont {W.}~\bibnamefont {Marshall}}, \bibinfo {author} {\bibfnamefont {T.}~\bibnamefont {Lindemann}}, \bibinfo {author} {\bibfnamefont {M.}~\bibnamefont {Manfra}}, \bibinfo {author} {\bibfnamefont {C.}~\bibnamefont {Marcus}},\ and\ \bibinfo {author} {\bibfnamefont {S.}~\bibnamefont {Vaitiek{\.e}nas}},\ }\bibfield  {title} {\bibinfo {title} {The hybrid josephson rhombus: A superconducting element with tailored current-phase relation},\ }\href@noop {} {\bibfield  {journal} {\bibinfo  {journal} {arXiv preprint arXiv:2406.20082}\ } (\bibinfo {year} {2024})}\BibitemShut {NoStop}%
\bibitem [{\citenamefont {Sun}\ \emph {et~al.}()\citenamefont {Sun}, \citenamefont {Feldstein-Bofill},\ and\ \citenamefont {et~al.}}]{zhenhaisun2025}%
  \BibitemOpen
  \bibfield  {author} {\bibinfo {author} {\bibfnamefont {Z.}~\bibnamefont {Sun}}, \bibinfo {author} {\bibfnamefont {D.}~\bibnamefont {Feldstein-Bofill}},\ and\ \bibinfo {author} {\bibnamefont {et~al.}},\ }\bibfield  {title} {\bibinfo {title} {(in preparation)},\ }\href@noop {} {\ }\BibitemShut {NoStop}%
\bibitem [{\citenamefont {Beenakker}(1991)}]{beenakker1991universal}%
  \BibitemOpen
  \bibfield  {author} {\bibinfo {author} {\bibfnamefont {C.}~\bibnamefont {Beenakker}},\ }\bibfield  {title} {\bibinfo {title} {Universal limit of critical-current fluctuations in mesoscopic josephson junctions},\ }\href@noop {} {\bibfield  {journal} {\bibinfo  {journal} {Physical review letters}\ }\textbf {\bibinfo {volume} {67}},\ \bibinfo {pages} {3836} (\bibinfo {year} {1991})}\BibitemShut {NoStop}%
\bibitem [{\citenamefont {Tinkham}(2004)}]{tinkham2004introduction}%
  \BibitemOpen
  \bibfield  {author} {\bibinfo {author} {\bibfnamefont {M.}~\bibnamefont {Tinkham}},\ }\href@noop {} {\emph {\bibinfo {title} {Introduction to superconductivity}}},\ Vol.~\bibinfo {volume} {1}\ (\bibinfo  {publisher} {Courier Corporation},\ \bibinfo {year} {2004})\BibitemShut {NoStop}%
\bibitem [{\citenamefont {Andersen}\ and\ \citenamefont {Blais}(2017)}]{andersen2017ultrastrong}%
  \BibitemOpen
  \bibfield  {author} {\bibinfo {author} {\bibfnamefont {C.~K.}\ \bibnamefont {Andersen}}\ and\ \bibinfo {author} {\bibfnamefont {A.}~\bibnamefont {Blais}},\ }\bibfield  {title} {\bibinfo {title} {Ultrastrong coupling dynamics with a transmon qubit},\ }\href@noop {} {\bibfield  {journal} {\bibinfo  {journal} {New Journal of Physics}\ }\textbf {\bibinfo {volume} {19}},\ \bibinfo {pages} {023022} (\bibinfo {year} {2017})}\BibitemShut {NoStop}%
\bibitem [{\citenamefont {Rower}\ \emph {et~al.}(2023)\citenamefont {Rower}, \citenamefont {Ateshian}, \citenamefont {Li}, \citenamefont {Hays}, \citenamefont {Bluvstein}, \citenamefont {Ding}, \citenamefont {Kannan}, \citenamefont {Almanakly}, \citenamefont {Braum{\"u}ller}, \citenamefont {Kim} \emph {et~al.}}]{rower2023evolution}%
  \BibitemOpen
  \bibfield  {author} {\bibinfo {author} {\bibfnamefont {D.~A.}\ \bibnamefont {Rower}}, \bibinfo {author} {\bibfnamefont {L.}~\bibnamefont {Ateshian}}, \bibinfo {author} {\bibfnamefont {L.~H.}\ \bibnamefont {Li}}, \bibinfo {author} {\bibfnamefont {M.}~\bibnamefont {Hays}}, \bibinfo {author} {\bibfnamefont {D.}~\bibnamefont {Bluvstein}}, \bibinfo {author} {\bibfnamefont {L.}~\bibnamefont {Ding}}, \bibinfo {author} {\bibfnamefont {B.}~\bibnamefont {Kannan}}, \bibinfo {author} {\bibfnamefont {A.}~\bibnamefont {Almanakly}}, \bibinfo {author} {\bibfnamefont {J.}~\bibnamefont {Braum{\"u}ller}}, \bibinfo {author} {\bibfnamefont {D.~K.}\ \bibnamefont {Kim}}, \emph {et~al.},\ }\bibfield  {title} {\bibinfo {title} {Evolution of 1/f flux noise in superconducting qubits with weak magnetic fields},\ }\href@noop {} {\bibfield  {journal} {\bibinfo  {journal} {Physical Review Letters}\ }\textbf {\bibinfo {volume} {130}},\ \bibinfo {pages} {220602} (\bibinfo {year} {2023})}\BibitemShut {NoStop}%
\bibitem [{\citenamefont {Azzalini}\ and\ \citenamefont {Valle}(1996)}]{azzalini1996multivariate}%
  \BibitemOpen
  \bibfield  {author} {\bibinfo {author} {\bibfnamefont {A.}~\bibnamefont {Azzalini}}\ and\ \bibinfo {author} {\bibfnamefont {A.~D.}\ \bibnamefont {Valle}},\ }\bibfield  {title} {\bibinfo {title} {The multivariate skew-normal distribution},\ }\href@noop {} {\bibfield  {journal} {\bibinfo  {journal} {Biometrika}\ }\textbf {\bibinfo {volume} {83}},\ \bibinfo {pages} {715} (\bibinfo {year} {1996})}\BibitemShut {NoStop}%
\bibitem [{\citenamefont {Schuster}\ \emph {et~al.}(2005)\citenamefont {Schuster}, \citenamefont {Wallraff}, \citenamefont {Blais}, \citenamefont {Frunzio}, \citenamefont {Huang}, \citenamefont {Majer}, \citenamefont {Girvin}, \citenamefont {Schoelkopf},\ and\ \citenamefont {RJ}}]{schuster2005ac}%
  \BibitemOpen
  \bibfield  {author} {\bibinfo {author} {\bibfnamefont {D.}~\bibnamefont {Schuster}}, \bibinfo {author} {\bibfnamefont {A.}~\bibnamefont {Wallraff}}, \bibinfo {author} {\bibfnamefont {A.}~\bibnamefont {Blais}}, \bibinfo {author} {\bibfnamefont {L.}~\bibnamefont {Frunzio}}, \bibinfo {author} {\bibfnamefont {R.~S.}\ \bibnamefont {Huang}}, \bibinfo {author} {\bibfnamefont {J.}~\bibnamefont {Majer}}, \bibinfo {author} {\bibfnamefont {S.}~\bibnamefont {Girvin}}, \bibinfo {author} {\bibnamefont {Schoelkopf}},\ and\ \bibinfo {author} {\bibnamefont {RJ}},\ }\bibfield  {title} {\bibinfo {title} {ac stark shift and dephasing of a superconducting qubit strongly coupled to a cavity field},\ }\href@noop {} {\bibfield  {journal} {\bibinfo  {journal} {Physical Review Letters}\ }\textbf {\bibinfo {volume} {94}},\ \bibinfo {pages} {123602} (\bibinfo {year} {2005})}\BibitemShut {NoStop}%
\bibitem [{\citenamefont {Luthi}\ \emph {et~al.}(2018)\citenamefont {Luthi}, \citenamefont {Stavenga}, \citenamefont {Enzing}, \citenamefont {Bruno}, \citenamefont {Dickel}, \citenamefont {Langford}, \citenamefont {Rol}, \citenamefont {Jespersen}, \citenamefont {Nyg{\aa}rd}, \citenamefont {Krogstrup} \emph {et~al.}}]{luthi2018evolution}%
  \BibitemOpen
  \bibfield  {author} {\bibinfo {author} {\bibfnamefont {F.}~\bibnamefont {Luthi}}, \bibinfo {author} {\bibfnamefont {T.}~\bibnamefont {Stavenga}}, \bibinfo {author} {\bibfnamefont {O.}~\bibnamefont {Enzing}}, \bibinfo {author} {\bibfnamefont {A.}~\bibnamefont {Bruno}}, \bibinfo {author} {\bibfnamefont {C.}~\bibnamefont {Dickel}}, \bibinfo {author} {\bibfnamefont {N.}~\bibnamefont {Langford}}, \bibinfo {author} {\bibfnamefont {M.~A.}\ \bibnamefont {Rol}}, \bibinfo {author} {\bibfnamefont {T.~S.}\ \bibnamefont {Jespersen}}, \bibinfo {author} {\bibfnamefont {J.}~\bibnamefont {Nyg{\aa}rd}}, \bibinfo {author} {\bibfnamefont {P.}~\bibnamefont {Krogstrup}}, \emph {et~al.},\ }\bibfield  {title} {\bibinfo {title} {Evolution of nanowire transmon qubits and their coherence in a magnetic field},\ }\href@noop {} {\bibfield  {journal} {\bibinfo  {journal} {Physical review letters}\ }\textbf {\bibinfo {volume} {120}},\ \bibinfo {pages} {100502} (\bibinfo {year} {2018})}\BibitemShut {NoStop}%
\bibitem [{\citenamefont {Casparis}\ \emph {et~al.}(2016)\citenamefont {Casparis}, \citenamefont {Larsen}, \citenamefont {Olsen}, \citenamefont {Kuemmeth}, \citenamefont {Krogstrup}, \citenamefont {Nyg{\aa}rd}, \citenamefont {Petersson},\ and\ \citenamefont {Marcus}}]{casparis2016gatemon}%
  \BibitemOpen
  \bibfield  {author} {\bibinfo {author} {\bibfnamefont {L.}~\bibnamefont {Casparis}}, \bibinfo {author} {\bibfnamefont {T.}~\bibnamefont {Larsen}}, \bibinfo {author} {\bibfnamefont {M.}~\bibnamefont {Olsen}}, \bibinfo {author} {\bibfnamefont {F.}~\bibnamefont {Kuemmeth}}, \bibinfo {author} {\bibfnamefont {P.}~\bibnamefont {Krogstrup}}, \bibinfo {author} {\bibfnamefont {J.}~\bibnamefont {Nyg{\aa}rd}}, \bibinfo {author} {\bibfnamefont {K.}~\bibnamefont {Petersson}},\ and\ \bibinfo {author} {\bibfnamefont {C.}~\bibnamefont {Marcus}},\ }\bibfield  {title} {\bibinfo {title} {Gatemon benchmarking and two-qubit operations},\ }\href@noop {} {\bibfield  {journal} {\bibinfo  {journal} {Physical review letters}\ }\textbf {\bibinfo {volume} {116}},\ \bibinfo {pages} {150505} (\bibinfo {year} {2016})}\BibitemShut {NoStop}%
\bibitem [{\citenamefont {Kringh{\o}j}\ \emph {et~al.}(2021)\citenamefont {Kringh{\o}j}, \citenamefont {Larsen}, \citenamefont {Erlandsson}, \citenamefont {Uilhoorn}, \citenamefont {Kroll}, \citenamefont {Hesselberg}, \citenamefont {McNeil}, \citenamefont {Krogstrup}, \citenamefont {Casparis}, \citenamefont {Marcus} \emph {et~al.}}]{kringhoj2021magnetic}%
  \BibitemOpen
  \bibfield  {author} {\bibinfo {author} {\bibfnamefont {A.}~\bibnamefont {Kringh{\o}j}}, \bibinfo {author} {\bibfnamefont {T.}~\bibnamefont {Larsen}}, \bibinfo {author} {\bibfnamefont {O.}~\bibnamefont {Erlandsson}}, \bibinfo {author} {\bibfnamefont {W.}~\bibnamefont {Uilhoorn}}, \bibinfo {author} {\bibfnamefont {J.}~\bibnamefont {Kroll}}, \bibinfo {author} {\bibfnamefont {M.}~\bibnamefont {Hesselberg}}, \bibinfo {author} {\bibfnamefont {R.}~\bibnamefont {McNeil}}, \bibinfo {author} {\bibfnamefont {P.}~\bibnamefont {Krogstrup}}, \bibinfo {author} {\bibfnamefont {L.}~\bibnamefont {Casparis}}, \bibinfo {author} {\bibfnamefont {C.}~\bibnamefont {Marcus}}, \emph {et~al.},\ }\bibfield  {title} {\bibinfo {title} {Magnetic-field-compatible superconducting transmon qubit},\ }\href@noop {} {\bibfield  {journal} {\bibinfo  {journal} {Physical Review Applied}\ }\textbf {\bibinfo {volume} {15}},\ \bibinfo {pages} {054001} (\bibinfo {year} {2021})}\BibitemShut {NoStop}%
\bibitem [{\citenamefont {Bargerbos}\ \emph {et~al.}(2023)\citenamefont {Bargerbos}, \citenamefont {Splitthoff}, \citenamefont {Pita-Vidal}, \citenamefont {Wesdorp}, \citenamefont {Liu}, \citenamefont {Krogstrup}, \citenamefont {Kouwenhoven}, \citenamefont {Andersen},\ and\ \citenamefont {Gr{\"u}nhaupt}}]{bargerbos2023mitigation}%
  \BibitemOpen
  \bibfield  {author} {\bibinfo {author} {\bibfnamefont {A.}~\bibnamefont {Bargerbos}}, \bibinfo {author} {\bibfnamefont {L.~J.}\ \bibnamefont {Splitthoff}}, \bibinfo {author} {\bibfnamefont {M.}~\bibnamefont {Pita-Vidal}}, \bibinfo {author} {\bibfnamefont {J.~J.}\ \bibnamefont {Wesdorp}}, \bibinfo {author} {\bibfnamefont {Y.}~\bibnamefont {Liu}}, \bibinfo {author} {\bibfnamefont {P.}~\bibnamefont {Krogstrup}}, \bibinfo {author} {\bibfnamefont {L.~P.}\ \bibnamefont {Kouwenhoven}}, \bibinfo {author} {\bibfnamefont {C.~K.}\ \bibnamefont {Andersen}},\ and\ \bibinfo {author} {\bibfnamefont {L.}~\bibnamefont {Gr{\"u}nhaupt}},\ }\bibfield  {title} {\bibinfo {title} {Mitigation of quasiparticle loss in superconducting qubits by phonon scattering},\ }\href@noop {} {\bibfield  {journal} {\bibinfo  {journal} {Physical Review Applied}\ }\textbf {\bibinfo {volume} {19}},\ \bibinfo {pages} {024014} (\bibinfo {year} {2023})}\BibitemShut {NoStop}%
\bibitem [{\citenamefont {Danilenko}\ \emph {et~al.}(2023)\citenamefont {Danilenko}, \citenamefont {Sabonis}, \citenamefont {Winkler}, \citenamefont {Erlandsson}, \citenamefont {Krogstrup},\ and\ \citenamefont {Marcus}}]{danilenko2023few}%
  \BibitemOpen
  \bibfield  {author} {\bibinfo {author} {\bibfnamefont {A.}~\bibnamefont {Danilenko}}, \bibinfo {author} {\bibfnamefont {D.}~\bibnamefont {Sabonis}}, \bibinfo {author} {\bibfnamefont {G.~W.}\ \bibnamefont {Winkler}}, \bibinfo {author} {\bibfnamefont {O.}~\bibnamefont {Erlandsson}}, \bibinfo {author} {\bibfnamefont {P.}~\bibnamefont {Krogstrup}},\ and\ \bibinfo {author} {\bibfnamefont {C.~M.}\ \bibnamefont {Marcus}},\ }\bibfield  {title} {\bibinfo {title} {Few-mode to mesoscopic junctions in gatemon qubits},\ }\href@noop {} {\bibfield  {journal} {\bibinfo  {journal} {Physical Review B}\ }\textbf {\bibinfo {volume} {108}},\ \bibinfo {pages} {L020505} (\bibinfo {year} {2023})}\BibitemShut {NoStop}%
\bibitem [{\citenamefont {Wang}\ \emph {et~al.}(2015)\citenamefont {Wang}, \citenamefont {Axline}, \citenamefont {Gao}, \citenamefont {Brecht}, \citenamefont {Chu}, \citenamefont {Frunzio}, \citenamefont {Devoret},\ and\ \citenamefont {Schoelkopf}}]{wang2015surface}%
  \BibitemOpen
  \bibfield  {author} {\bibinfo {author} {\bibfnamefont {C.}~\bibnamefont {Wang}}, \bibinfo {author} {\bibfnamefont {C.}~\bibnamefont {Axline}}, \bibinfo {author} {\bibfnamefont {Y.~Y.}\ \bibnamefont {Gao}}, \bibinfo {author} {\bibfnamefont {T.}~\bibnamefont {Brecht}}, \bibinfo {author} {\bibfnamefont {Y.}~\bibnamefont {Chu}}, \bibinfo {author} {\bibfnamefont {L.}~\bibnamefont {Frunzio}}, \bibinfo {author} {\bibfnamefont {M.}~\bibnamefont {Devoret}},\ and\ \bibinfo {author} {\bibfnamefont {R.~J.}\ \bibnamefont {Schoelkopf}},\ }\bibfield  {title} {\bibinfo {title} {Surface participation and dielectric loss in superconducting qubits},\ }\href@noop {} {\bibfield  {journal} {\bibinfo  {journal} {Applied Physics Letters}\ }\textbf {\bibinfo {volume} {107}} (\bibinfo {year} {2015})}\BibitemShut {NoStop}%
\end{thebibliography}%

\end{document}